\documentclass{appolb}
\Preprinttrue
\usepackage{mathbbol}
\def\beq{\begin{equation}}
\def\eeq{\end{equation}}
\def\bea{\begin{array}}
\def\eea{\end{array}}
\def\beqa{\begin{eqnarray}}
\def\eeqa{\end{eqnarray}}
\def\Pexp{{\rm Pexp}}
\def\cO{{\cal{O}}}
\def\cP{{\cal{P}}}
\def\cA{{\cal{A}}}

\def\cF{{\cal{F}}}
\def\cV{{\cal{V}}}
\def\cR{{\cal{R}}}
\def\cD{{\cal{D}}}
\def\k{{\bf k}}
\def\ff{{\rm{ff}}}
\def\zm{{\rm{zm}}}
\def\ddz{{\frac{d}{dz}}}
\def\Tr{{\rm Tr}}
\def\tr{{\rm tr}}
\def\diag{{\rm diag}}
\def\pl{{{\cal P}_\infty}}

\newcommand{\re}{\relax{\rm I\kern-.18em R}}

\newcommand{\refeq}[1]{\mbox{Eq.~(\ref{eq:#1})}}
\renewcommand{\half}{{\scriptstyle{{1\over 2}}}}
\newcommand{\hhalf}{{\scriptstyle{{1/2}}}}

\newcommand{\twthd}{{\scriptstyle{{2\over 3}}}}

\newcommand{\veps}{\varepsilon}

\newcommand{\zahlen}{{\mathbb{Z}}}
\newcommand{\ein}{{\Eins}}
\newcommand{\inv}{{-1}}

\begin{document}
\title{INSTANTONS AND CONSTITUENT MONOPOLES
\thanks{Presented by the last author at the Cracow School of 
Theoretical Physics, XLIII Course, Zakopane, May 30 - June 8, 
2003}
\vskip-3cm\hfill INLO-PUB-12/03\vskip2.5cm
}
\author{Falk Bruckmann, D\'aniel N\'ogr\'adi and 
Pierre van Baal
\address{Instituut-Lorentz for Theoretical Physics, University
of Leiden,\\ P.O.Box 9506, NL-2300 RA Leiden, The Netherlands}
\vskip5mm
\address{\Large Dedicated to the memory of Ian Kogan}
\vskip-2mm
}
\maketitle
\begin{abstract}
We review how instanton solutions at finite temperature can be seen 
as boundstates of constituent monopoles, discuss some speculations 
concerning their physical relevance and the lattice evidence for 
their presence in a dynamical context.
\end{abstract}
  
\section{Introduction}
Over the last five years there has been a revived interest in studying
instantons at finite temperature $T$, so-called calorons~\cite{HaSh,GPY}. 
The main reason is that new explicit solutions could be obtained in the 
case where the Polyakov loop at spatial infinity is non-trivial, necessary 
to reveal more clearly the constituent nature of these calorons. This 
asymptotic value of the Polyakov loop is called the holonomy. In gauge 
theories trivial holonomy, for which the asymptotic value of the Polyakov 
loop takes values in the center of the gauge group, is typical for the 
deconfined phase. Therefore, caloron solutions with non-trivial holonomy 
are more expected to play a role in the confined phase, still at finite 
temperature, but where the average of the trace of the Polyakov loop is 
small. In the introduction we start with a pedagogical overview discussing 
monopoles, instantons, and their physical significance. The next section 
discusses the construction of the caloron solutions. Secs.~\ref{sec:adhm} and 
\ref{sec:nahm} are more technical, but tutorial in nature. One could skip 
in particular Secs.~\ref{sec:expl} and \ref{sec:spec}, describing in 
Sec.~\ref{sec:prop} the properties of the solutions in less technical 
terms. Lattice evidence for the dynamical significance of constituent 
monopoles is discussed in Sec.~\ref{sec:latt}.

\subsection{Monopoles}\label{sec:mono}
That monopoles should play a role in describing the constituent nature 
of calorons is in itself not really a surprise, because at finite 
temperature $A_0$ plays in some sense the role of a Higgs field in the 
adjoint representation. However, a gauge transformation,
\beq
{}^gA_\mu(x)=g(x)A_\mu(x)g^\inv(x)+g(x)\partial_\mu g^\inv(x),
\eeq
shows that $A_0$ does not transform correctly (unless the gauge 
transformation is time independent), due to the inhomogeneous term. 
Instead, the Polyakov loop 
\beq
P(t,\vec x)=\Pexp\left(\int_0^\beta A_0(t+s,\vec x)ds\right),
\eeq
transforms as it should, ${}^gP(x)=g(x)P(x)g^\inv(x)$. Here $\beta=1/kT$
is the period in the imaginary time direction, under which the gauge field 
is assumed to be periodic. We also will consider other gauges, where 
$A_\mu(x)$ is periodic up to a gauge transformation, in which case the 
expression for $P$ has to be modified accordingly~\cite{Iof}. For example, 
in the so-called algebraic gauge, $A_0(x)$ is transformed to zero at 
spatial infinity. In this case the gauge fields satisfy the boundary 
condition ($\pl$ to be defined below)
\beq
A_\mu(t+\beta,\vec x)=\pl A_\mu(t,\vec x)\cP_\infty^\inv.\label{eq:agauge}
\eeq

We will require that the total Euclidean action of these calorons
is finite, such that the field strength\footnote{Our conventions are 
$A_\mu(x)=e A_\mu^a(x)T_a$, where $e$ is the coupling constant and
$T_a^\dagger=-T_a$, $\Tr(T_aT_b)=-\half\delta_{ab}$, $[T_a,T_b]=
f_{abc}T_c$. For SU(2) $f_{abc}=\veps_{abc}$ and $T_a=-i\half\tau_a$ 
in terms of the familiar Pauli matrices.}
\beq
F_{\mu\nu}(x)=\partial_\mu A_\nu(x)-\partial_\nu A_\mu(x)+
[A_\mu(x),A_\nu(x)]
\eeq
has to go to zero at spatial infinity. It is this that forces
the Polyakov loop to become constant at spatial infinity. For
SU($n$) gauge theory this gives
\beq
\pl\!=\!\exp(\beta A_0^\infty)\!\equiv\!\lim_{|\vec x|\to\infty}\!\!P(x),\quad 
A_0^\infty\!\equiv\!\frac{2\pi i}{\beta}U_0\diag\left(\mu_1,\mu_2,\ldots,
\mu_n\right)U_0^\inv\!,\label{eq:pol}
\eeq
independent of the direction and time. Unlike for a Higgs field, however, 
$\pl$ is unitary with determinant 1. Choosing $U_0$ (the constant gauge 
function that brings $\pl$ to its diagonal form) appropriately, the $n$ 
eigenvalues, $\exp(2\pi i\mu_m)$, are ordered. The freedom in adding an 
arbitrary integer to $\mu_m$ is fixed by ordering the $\mu_m$ themselves, 
and requiring their sum to vanish,
\beq
\mbox{$\sum_{i=1}^n\mu_i=0$},\quad
\mu_1\leq\mu_2\leq\ldots\leq\mu_n\leq\mu_{n+1}\equiv1+\mu_1.\label{eq:order}
\eeq
For trivial holonomy $\pl$ is an element of the center of the gauge 
group, $\pl=\exp(2\pi iq/n)\Eins_n$ with $q$ an integer between 
0 and $n-1$, hence $\mu_m=q/n$.

One might now immediately object that a Higgs field that goes to a constant 
at infinity does not have the usual hedgehog form expected for a non-Abelian 
't~Hooft-Polyakov monopole~\cite{Mon}, but this is because the caloron 
solutions are actually such that the total magnetic charge vanishes. The 
force stability of these solutions is based, as for exact multi-monopole 
solutions in the Bogomol'ny-Prasad-Sommerfeld (BPS) limit~\cite{BPS} on 
balancing the electromagnetic with the scalar (Higgs) force~\cite{Mant,MoOl}. 
For the caloron the difference is in interchanging repulsive and attractive 
forces. For a single caloron with topological charge one, this is because 
there are $n-1$ monopoles with a unit magnetic charge in the $i$-th U(1) 
subgroup, all compensated by the $n$-th monopole of so-called type $(1,1,
\ldots,1)$, having a magnetic charge in each of these subgroups. This special 
monopole is also called a Kaluza-Klein (KK) monopole~\cite{LeYi}, as will be 
explained below. The well-known Harrington-Shepard solution~\cite{HaSh} has 
trivial holonomy, and although all eigenvalues of $\pl$ are equal, such that 
there is no spontaneous symmetry breaking, one still finds a genuine BPS 
monopole~\cite{Ros} in a suitable limit (it will be the KK monopole that 
survives).

In a Higgs theory, switching off the Higgs potential and splitting off a 
square in the energy density, 
\beq
-\Tr\left((D_i\Phi)^2+B_i^2\right)=-\Tr(D_i\Phi\mp B_i)^2\mp2
\Tr\left(B_iD_i\Phi\right),
\eeq
{\em exact} monopole solutions are constructed using the BPS 
condition~\cite{BPS}, which imposes the covariant derivative of the Higgs 
field $\Phi$ to be equal (up to a sign) to the magnetic field, $D_i\Phi\equiv
\partial_i\Phi+[A_i,\Phi]=\pm B_i$, where $B_i\equiv\half\veps_{ijk}F_{jk}$. 
One is then left with a total derivative 
\beq
\Tr\left(B_iD_i\Phi\right)=\Tr\left(D_i(B_i\Phi)\right)=\partial_i\Tr(B_i\Phi),
\eeq
whose integral is proportional to the magnetic charge. The integral can also 
be associated to the mapping degree of the map $\hat x\to\Phi(r\hat x)$ ($\hat 
x\equiv\vec x/|\vec x|$) for $r\to\infty$. With $\Phi$ taking values in the 
algebra, and of fixed length at infinity, this gives for SU(2) a map from 
$S^2$ to $S^2$. For a caloron the BPS condition is simply a consequence of 
the self-duality conditions characteristic of instanton solutions, $E_i=D_iA_0
-\partial_0A_i=\pm B_i$, with $\Phi=A_0$. One might thus be tempted to call 
the constituents dyons, rather than monopoles. In the Higgs model the 
Julia-Zee dyons are constructed by taking $A_0$ proportional to the Higgs 
field $\Phi$~\cite{JuZe}. By a time dependent gauge transformation 
$A_0$ can then be gauged to zero. The resulting electric field is now given 
by $E_i=-\partial_0A_i$ and is {\em not} quantized (and in particular 
not equal to $\pm B_i$). In pure gauge theory it makes, however, no sense 
to separate $D_i\Phi=D_iA_0$ from $\partial_0A_i$. Gauge invariance requires 
that they occur in the combination $F_{i0}=D_iA_0-\partial_0A_i$. The 
electric field is necessarily fixed and quantized as soon as we interpret 
$A_0$ as the Higgs field. As discussed in Ref.~\cite{KvB2} adding a fifth 
dimension for Minkowski time, compactifying the Euclidean time direction 
to zero size ($\beta\to0$) allows one to have electric charge as for the 
Julia-Zee dyon. A compactified Euclidean direction is what one also considers
in Kaluza-Klein theories. It is in this sense that the type $(1,1,\ldots,1)$
monopole is called a KK monopole, because it turns out it is static up to a 
gauge transformation that makes one full rotation in the unbroken subgroup 
when going from 0 to $\beta$, thus of lowest non-trivial Kaluza-Klein 
momentum\footnote{Nevertheless, there is a context in pure gauge theories 
where dyons appear, but for this we have to add a term proportional to 
$\theta F\tilde F$ to the Lagrangian~\cite{Wit}. The electric charge is 
now given by $\theta/(2\pi)$ times the magnetic charge.}.

\subsection{Instantons}\label{sec:inst}
\begin{figure}[htb]
\vspace{3.3cm}
\includegraphics{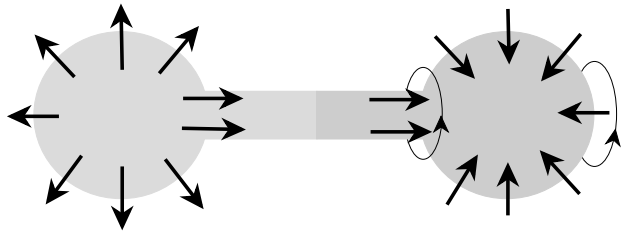}
\caption{The topologically non-trivial field configuration is constructed 
from two oppositely charged monopoles, rotating one of them over a full 
$2\pi$ rotation.}\label{fig:1}
\end{figure}

We therefore consider it most appropriate to call the constituents mono\-poles, 
as is also clear from Nahm's formalism which provides one of the essential 
tools to find these self-dual solutions~\cite{Nahm}. The Nahm transformation, 
as well as the Atiyah-Drinfeld-Hitchin-Manin (ADHM) construction~\cite{ADHM} 
for instanton solutions on $\re^4$, form indispensable tools to find the exact 
caloron solutions. It should be added that from the topological point of view 
Taubes showed how to make out of two oppositely charged monopoles a Euclidean 
four dimensional gauge field that has non-zero topological charge~\cite{Taub}. 
This result is more general, since only when minimizing the action in a sector 
with non-trivial topological charge will one find a self-dual instanton 
solution. His construction is based on creating a monopole anti-monopole pair, 
bringing them far apart, rotating one of them over a full rotation (the 
so-called Taubes winding) and finally bringing them together to annihilate, 
see Fig.~\ref{fig:1}. The four dimensional configuration constructed this 
way is topologically non-trivial.

In $\re^4$ the topological charge is related to the winding of a gauge 
function $g(x)$ defined at infinity, which therefore is a mapping from 
$S^3$ to the gauge group. The way $g(x)$ enters is through requiring 
the Euclidean action, 
\beq
S=-\frac{1}{2e^2}\int d^4x~\Tr\left(F^2_{\mu\nu}(x)\right),
\eeq
to remain finite. As before this implies the field strength at infinity to
go to zero, where $A_\mu(x)$ can be written as a pure gauge, $A_\mu(x)=g(x)
\partial_\mu g^\inv(x)$. For $SU(2)$ a simple parametrization as $g(x)=a_\mu(x)
\sigma_\mu$ in terms of a unit vector $a_\mu$ and unit quaternions $\sigma_\mu
=(\Eins_2,i\vec\tau)$ ($\bar\sigma_\mu\equiv\sigma_\mu^\dagger$) makes this 
winding most transparent as the mapping degree of $a(x)$ at infinity. In the 
simplest case of degree 1, $a(x)=x/|x|$ is a 4-dimensional hedgehog. The 
relation to Taubes' construction is using the fact that $S^3$ can be viewed 
as a twisted product of $S^1$ (the Taubes winding) and $S^2$ (the hedgehog 
formed by the Higgs field), the so-called Hopf fibration~\cite{KvB2,KvBB}. 

In the case of periodicity in the imaginary time direction as it occurs for 
calorons, let us transform $A_0(x)\to0$ everywhere. This can be done by a 
time dependent gauge transformation $U(x)$, which in general will {\em not} 
be periodic. Actually $P(t;\vec x)$ itself is the gauge transformation that 
relates the gauge field at $t+\beta$ to that at $t$ in the $A_0=0$ gauge. 
Since $P(x)$ goes to a constant ($\pl$) at spatial infinity, this provides 
a non-trivial mapping from $S^3$ (as $\re^3$ compactified at infinity) to the 
gauge group. The topological charge is precisely its winding number. 

As in the case of the monopole energy density, we can rewrite the action 
density in terms of a square and a boundary term, 
\beqa
\Tr(F_{\mu\nu}^2)&=&\half\Tr(F_{\mu\nu}\pm\tilde F_{\mu\nu})^2\mp
\Tr(F_{\mu\nu}\tilde F_{\mu\nu}),\quad\Tr(F_{\mu\nu}\tilde F_{\mu\nu})
=\partial_\mu K_\mu,\nonumber\\K_\mu&=&2\veps_{\mu\nu\alpha\beta}
\Tr(A_\nu\partial_\alpha A_\beta+\twthd A_\nu A_\alpha A_\beta),
\eeqa
with $\tilde F_{\mu\nu}=\half\veps_{\mu\nu\alpha\beta}F_{\alpha\beta}$ 
the dual field strength interchanging electric and magnetic components. 
Hence, for self-dual solutions at finite temperature 
\beq
S=\frac{1}{2e^2}\int\left(K_0(0,\vec x)-K_0(\beta,\vec x)
\right)d^3x=\frac{1}{3e^2}\int\Tr(P(\vec x)dP^\inv(\vec x))^3,
\eeq
using that in the $A_0=0$ gauge $A(\beta,\vec x)={}^PA(0,\vec x)$, 
expressing the result in a compact differential form notation.
Note that the integral is invariant under small deformations of $P$, using 
$\delta(PdP^\inv)^3=3d\Tr(P\delta P^\inv(PdP^\inv)^2)$, and therefore should
be proportional to the winding number of $P$. A map of degree $k$ can be 
obtained by multiplying $k$ maps of degree 1 (the inverse gives a negative 
mapping degree). Each of these maps of degree 1 can be deformed at will, 
as the integral is invariant under continuous deformations, and in 
particular can be arranged to be the identity except for a small region. 
Choosing these regions to have no overlap, the integral is easily seen to be 
proportional to $k$. To fix for SU(2) the constant of proportionality we may 
take $U(\vec x)=((1-|\vec x|^2)\sigma_0+2\vec x\cdot\vec\sigma)/(1+|\vec 
x|^2)$, for the map of degree 1 related to stereographic projection from 
$S^3$ to $\re^3$. For SU($n$) we first deform the map to lie in an SU(2) 
subgroup. This then gives the celebrated result that self-dual solutions 
with given topological charge, $k=(16\pi^2)^{-1}\int d^4x\Tr\left(F_{\mu\nu}
(x)\tilde F_{\mu\nu}(x)\right)$, have an action equal $8\pi^2|k|/e^2$. 

This is also a convenient setting to understand why in the limit of zero 
temperature, $\beta\to\infty$, an instanton corresponds to vacuum to 
vacuum tunneling. Finite action requires the field strength to go to 
zero at $|t|\to\infty$, but at the same time, the field at $t\to\infty$ 
is related to the field at $t\to-\infty$ by a topologically non-trivial 
gauge transformation. Strictly speaking, we need to identify gauge field 
configuration that are gauge equivalent. Instead of having multiple vacua 
we can alternatively say there is one vacuum, with the field space being
non-contractible. This is quite analogous to considering a periodic 
quantum mechanics problem, which could be reinterpreted as quantum 
mechanics on a circle. Tunneling now corresponds to penetration of the 
wave function in the classically forbidden region, to reach back to the 
vacuum going around the circle in either direction. Essential is that
the support of the wave function, the region where it is non-vanishing,
becomes sensitive to the non-trivial topology of the configuration 
space~\cite{Iof}. 

\begin{figure}[htb]
\vspace{4cm}
\includegraphics{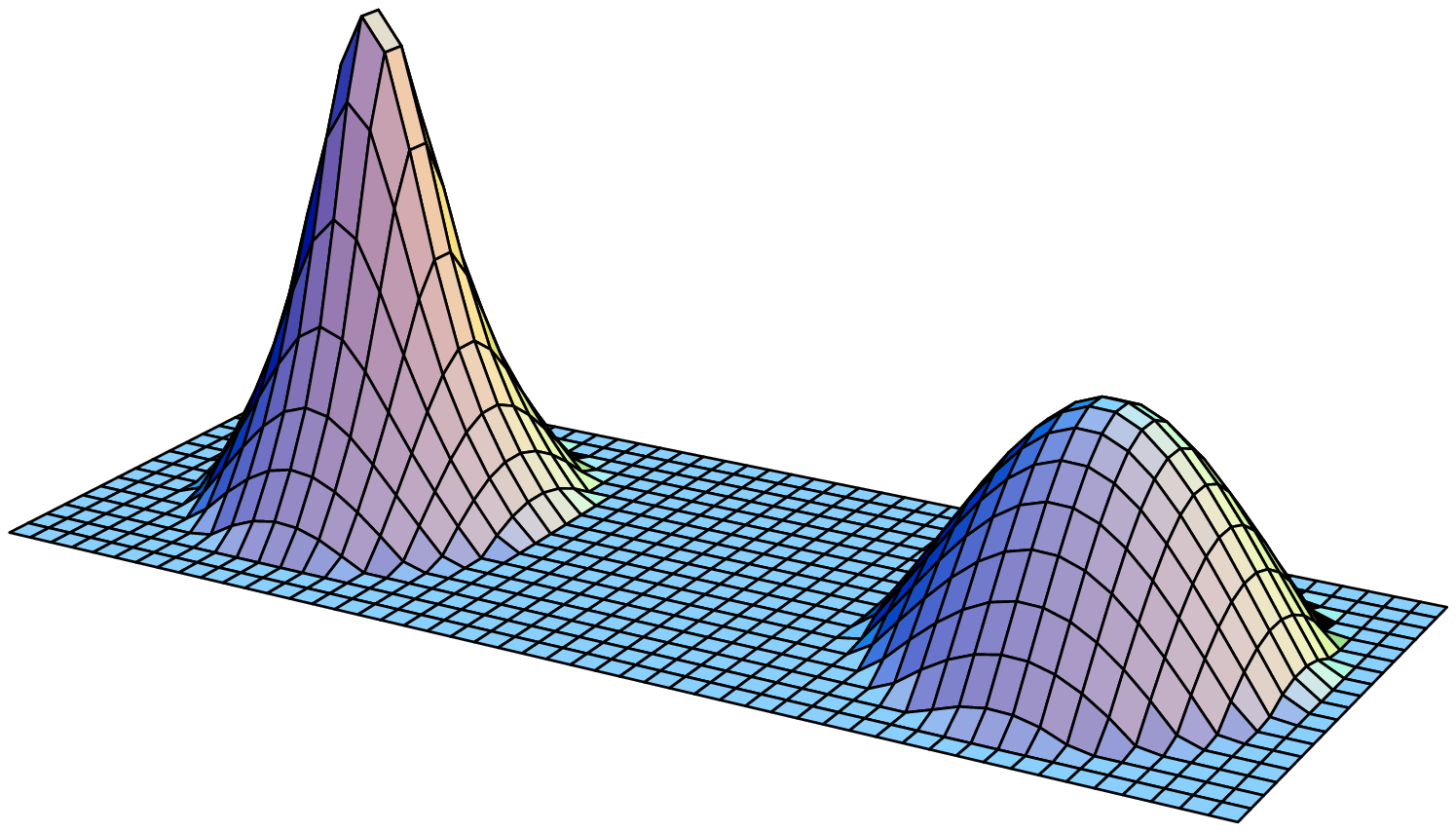}
\includegraphics{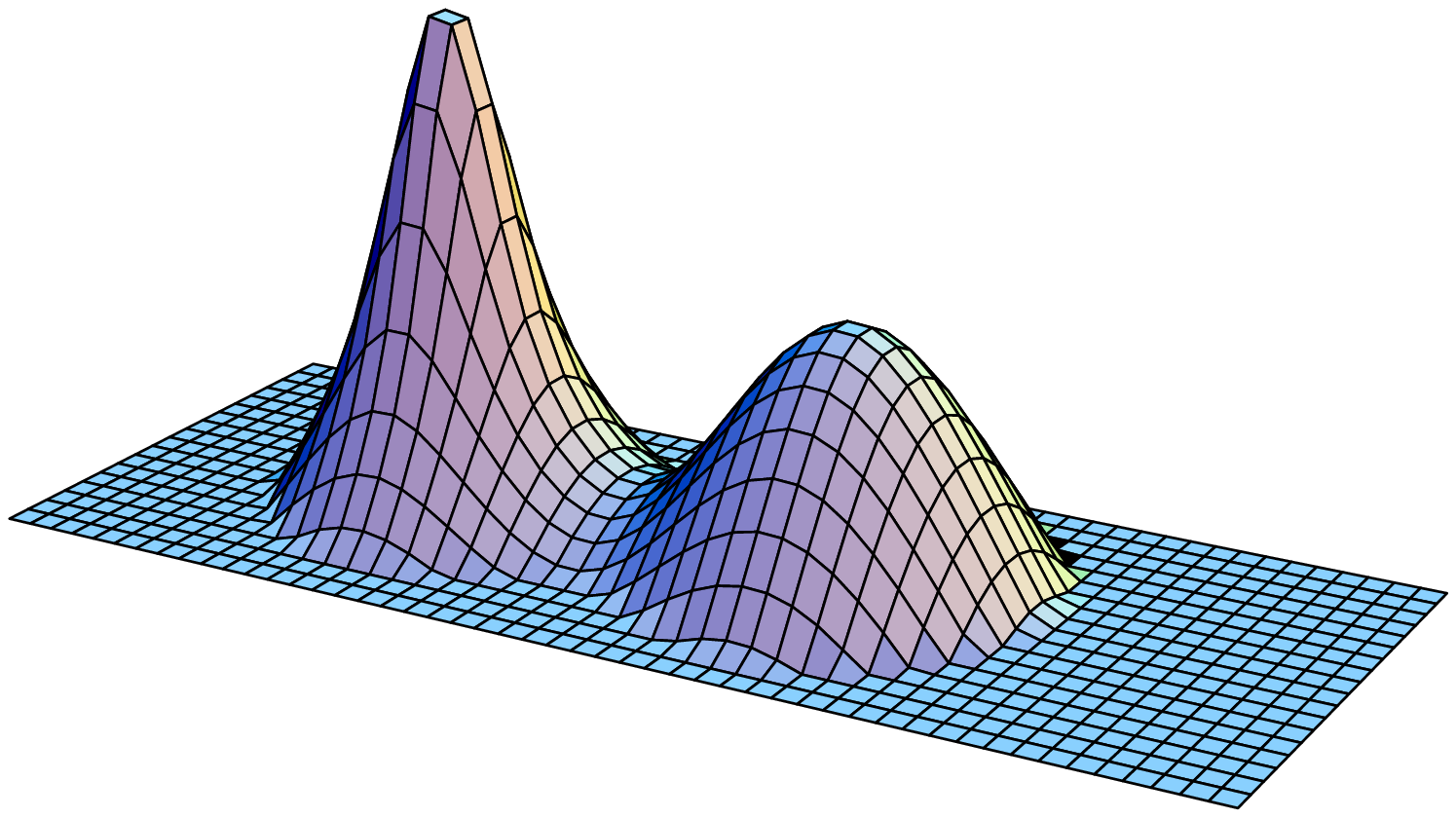}
\includegraphics{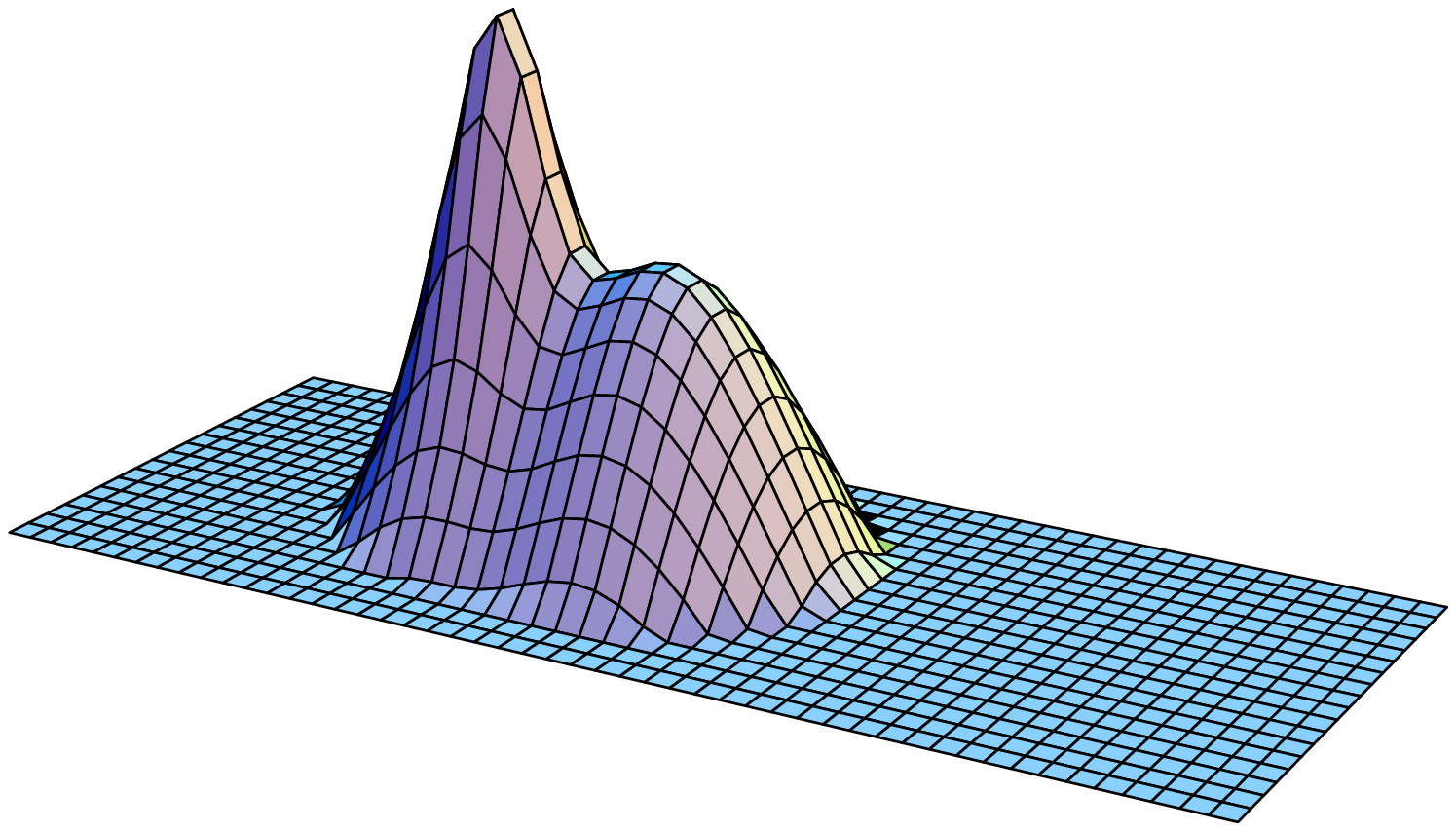}
\caption{Shown is a typical example for the action density (on equal
logarithmic scales at the time where it is maximal) of an SU(2) caloron 
with non-trivial holonomy $\mu_2=-\mu_1=0.125$ (for, from left to right, 
$\rho/\beta=0.8,1.2,1.6$).}\label{fig:2}
\end{figure}

The tunneling path in non-Abelian gauge theories is of course described 
by the one parameter ($t$) family of gauge fields $A_i(t,\vec x)$ (in the 
$A_0=0$ gauge), and a finite action means we have to cross a potential 
barrier when going around the non-contractible loop (or going from 
vacuum to vacuum). From Taubes' argument it is already clear that this 
intermediate configuration can be associated to a monopole-antimonopole 
boundstate, which is made more precise by the caloron solutions to be
described. At zero-temperature this is less obvious, since there monopoles 
form close pairs, cmp. Fig.~\ref{fig:2} (left). They behave very much like 
virtual particles, that can only be created for a short period of time, 
thereby explaining the instantaneous character. At finite temperature, the 
monopoles can be separated much more easily due to the interactions with the 
thermal bath. At high temperature, however, they will be suppressed due to the 
Boltzmann weight, and so monopoles (like calorons) are expected to be dilute. 
At low temperature instantons form a more dense ensemble, possibly leading
to monopoles to be dense as well, particularly when instantons overlap.
One may then have some hope that the confining electric phase could be 
characterized by a deconfining magnetic phase, where the dual deconfinement 
is due to the large monopole density, in a similar spirit to high density 
induced quark deconfinement. It would offer a possible alternative to 
existing scenarios.

Instantons describe virtual processes and are quite often discussed in 
the context of the semiclassical approximation. For a one-dimensional 
particle with mass $m$ in a positive potential $V$ we may again split
off a square,
\beqa
&&\hskip-5mm L(t)\!=\!\half m\dot x^2(t)\!+\!V(x(t))\!=\!\half m\left(\dot 
x(t)-\sqrt{2V(x(t))/m}\right)^2\!\!+\dot x(t)\sqrt{2mV(x(t))},\nonumber\\
&&\hskip-3.5mm S_{cl}=\int dt~\dot x(t)\sqrt{2mV(x(t))}=\int dx\sqrt{2mV(x)},
\eeqa
where we typically integrate between the classical turning points, related 
to the WKB expression for the wave function in the classically forbidden 
region, $\exp(-\int_{x_0}^x dy\sqrt{2mV(y)}/\hbar)$. In a double well this 
leads to a tunnel splitting proportional to $\exp(-S_{cl}/\hbar)$. 
One problem in the theory of strong interactions is that the effective 
coupling tends to become too big for large instantons. Instantons can 
have an arbitrary size $\rho$, due to the classical scale invariance of 
the theory. This classical scale invariance gets broken by the 
regularization and one is left with a scale dependent running coupling 
constant after renormalization. This causes a problem when integrating over 
the scale parameter in the one-instanton tunneling amplitude given by the 
celebrated result of 't~Hooft~\cite{Hoo}, $dW\propto d\rho dx^4\rho^{-5}\exp
\left(-8\pi^2/e^2(\rho)\right)$. Actually, for calorons with non-trivial 
holonomy and $\rho>\beta$, it is more natural to associate $\rho$ with 
the distance between constituent monopoles ($\pi\rho^2/\beta$ for SU(2)), 
and this may help alleviate the problem one encounters when dealing with 
large scale instantons. This is also one way to understand why there will 
still be calorons with $\rho$ arbitrary, despite the fact that $\beta$ 
fixes the scale\footnote{Remarkably, it can be shown~\cite{Taub2,DoKr} under 
some mild conditions, that any four dimensional manifold will have instanton 
solutions with arbitrary $\rho$.}. At zero temperature, a large $\rho$ leads 
to a low energy barrier along the tunneling path, and at some point this will 
no longer describe a virtual process and the semiclassical approximation will 
break down. Nevertheless, the instanton liquid model has been very successful 
in describing much of the low energy phenomenology, in particular for chiral 
dynamics and aspects related to breaking the axial U(1) symmetry. We refer 
to the reviews by Sch\"afer and Shuryak~\cite{ScSh} and by Diakonov~\cite{Diak} 
for more details. 

\subsection{Fermion zero-modes}\label{sec:ferm}
\begin{figure}[htb]
\vskip1.7cm\hskip1.8cm$E=0$\vskip2.3cm
\includegraphics{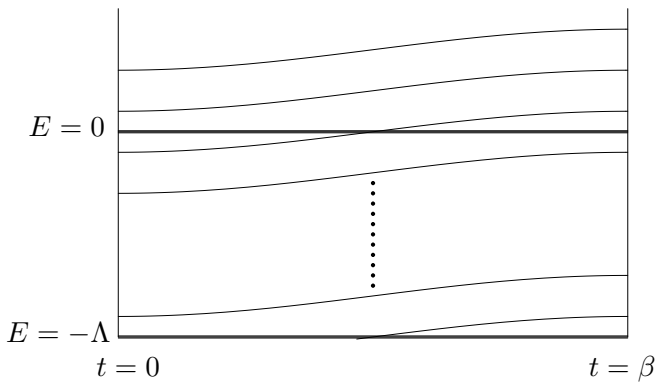}
\vskip0mm\hskip1.5cm$E=-\Lambda$\vskip0mm
\hskip2.7cm$t=0$\hskip5.7cm$t=\beta$
\caption{Schematic representation of the spectral flow.}\label{fig:3}
\end{figure}

In the instanton liquid model the interaction of instantons with fermions 
plays an important role. This is because instantons have a remarkable 
influence on the spectrum of Dirac fermions, it namely implies the presence 
of zero eigenvalue solutions with fixed chirality (the so-called chiral 
zero-modes). These chiral zero-modes will play an important role in the Nahm 
transformation and in studying the properties of the calorons with non-trivial 
holonomy, so we wish to mention some of the interesting and far reaching 
physical consequences. This is most easily discussed in terms of the spectral 
flow of the Dirac Hamiltonian along the tunneling path. Its spectrum is of 
course gauge invariant, and this implies that the energy levels at $t=0$ and 
$t=\beta$ are identical. The gauge field provides a smooth interpolation 
between these, which leads to each of the energies to be a continuous 
function of $t$, the so-called spectral flow, see Fig~\ref{fig:3}. 
If we draw the energy level at zero, the number of crossings is clearly a 
conserved quantity, it cannot change under continuous deformations of the gauge
field background (the background need not be a solution of the equations of 
motion). It is also clear, by putting instantons in a row, that the number of 
crossings is proportional to the topological charge, and not surprisingly it 
is actually equal to it. The necessity of such a crossing is related to the 
existence of a chiral zero-mode, whose number for fermions in the fundamental 
representation is equal to the topological charge, as follows from the 
Atiyah-Singer index theorem~\cite{AtSi}. The argument is most simple in the 
case of zero-temperature, where the zero-mode at $t\to\pm\infty$ has to behave 
as $\exp(-tE_\pm)$, where $E_\pm$ is one of (free) fermion energies. Clearly 
this would only give a normalizable zero-mode if $E_+>0$ and $E_-<0$, which 
forces a crossing! At this point the Dirac vacuum is degenerate and particles 
or anti-particles are created with definite chirality. Compatible with the 
breaking of the axial U(1) symmetry, the divergences of its current is 
proportional to the topological density $\tilde F_{\mu\nu}F_{\mu\nu}$. It 
gives rise to the so-called 't~Hooft interaction~\cite{Hoo}, which together 
with a finite density of fermion modes at zero eigenvalue (required for the 
Banks-Casher mechanism~\cite{BaCa} to work) play an important role in the 
spontaneous breaking of chiral symmetry (or soft breaking in case of small 
up and down quark masses)~\cite{ScSh}. Finally, the spectral flow also makes 
it easy to understand the origin of the axial anomaly, which occurs due to 
the need to regularize the theory.  If cutting off modes with $E<-\Lambda$, 
where $\Lambda$ is the ultraviolet cutoff, we see that the spectral flow 
leads to the fact that modes we had removed in the trivial vacuum reappear 
due to the spectral flow at the vacuum equivalent to it by a topologically 
non-trivial gauge transformation. That the violation in conserving the axial 
U(1) charge is proportional to the topological charge is in this setting 
simply a consequence of the number of crossings in the spectral flow. This 
is the celebrated infrared-ultraviolet connection, and also makes it 
understandable why the anomaly is robust (fully determined by the lowest 
order result in perturbation theory).

\section{Construction of solutions}\label{sec:cons}
Let us start with the well know SU(2) Harrington-Shepard solution~\cite{HaSh}
for the caloron with trivial holonomy. In that case one simply takes a periodic
array of instantons parallel in group space ($\pl=\Eins_2$), placed at 
$a_{(p)}\equiv(a_0+p\beta,\vec a)$ for integer $p$. The solution is 
found in terms of the 't~Hooft ansatz~\cite{Raj}, which in its general form 
is given by
\beq
A_\mu(x)=\half\bar\eta_{\mu\nu}\partial_\nu\log\phi(x).\label{eq:tHooft}
\eeq
Here $\eta_{\mu\nu}=\half(\sigma_\mu\bar\sigma_\nu-\sigma_\nu\bar\sigma_\mu)=
i\eta_{\mu\nu}^a\tau_a$ and $\bar\eta_{\mu\nu}=\half(\bar\sigma_\mu\sigma_\nu
-\bar\sigma_\nu\sigma_\mu)=i\bar\eta_{\mu\nu}^a\tau_a$, are the self-dual
and anti-selfdual 't~Hooft tensors. Substituting this in the self-duality 
equation, $F_{\mu\nu}=\tilde F_{\mu\nu}$, one finds this is a solutions 
if and only if $\partial_\mu^2\phi(x)=0$. Therefore $\phi=1+\sum_{i=1}^k
\rho_i^2|x-b_{(i)}|^2$, where $b_{(i)}$ are the four dimensional locations 
and $\rho_i$ the sizes of the $k$ instantons\footnote{Using conformal 
transformations a generalization of this ansatz including non-trivial 
color orientations exists~\cite{JNR}, but it will only give all possible 
solutions for charge 2.}. A singularity at $x=b_{(i)}$ can be removed by 
a gauge transformation. For the Harrington-Shepard solution, taking 
$b_{(p)}=a_{(p)}$ and $\rho_p=\rho$ one can perform the sum over 
$p\in\zahlen$ to find~\cite{HaSh},
\beq
\phi_{HS}(x)=1+\frac{\pi\rho^2\sinh(2\pi r/\beta)/(\beta r)}{\cosh(2\pi r/
\beta)-\cos(2\pi(t-a_0)/\beta)},\quad r=|\vec x-\vec a|.\label{eq:HS}
\eeq

The case of non-trivial holonomy cannot be treated in the same way,
because we will have to sum over a periodic array of instantons that
has a color rotation by $\pl$ when shifting time over $\beta$. This
can only be treated within the full ADHM ansatz. The rest of this 
section is more technical and could be skipped, although the short 
tutorials on the ADHM construction and the Nahm transformation in 
Secs.~\ref{sec:adhm} and \ref{sec:nahm} are still recommended.

\subsection{ADHM formalism}\label{sec:adhm}
The $SU(n)$ ADHM construction for charge $k$ instantons~\cite{ADHM} starts 
with a $k$ dimensional vector $\lambda=(\lambda_1,\ldots,\lambda_k)$, where 
$\lambda_i^\dagger$ is a two-component spinor in the $\bar n$ representation 
of $SU(n)$ (i.e. $\lambda$ is a $n\times 2k$ complex matrix) and a $2k\times
2k$ complex matrix $B=\sigma_\mu\otimes B_\mu$ (each $B_\mu$ is a hermitian 
$k\times k$ matrix). These are combined to form a $(n + 2 k)\times 2k$ 
dimensional matrix $\Delta(x)$, which has $n$ normalized eigenvectors 
with vanishing eigenvalue, combined in a complex matrix $v(x)$ of size 
$(n+2k)\times n$,
\beq
\Delta(x)=\left(\!\!\bea{c}\lambda\\B(x)\eea\!\!\right),\quad B(x)=B-x\Eins_k,
\quad\Delta^\dagger(x)v(x)=0.
\eeq
Here the quaternion $x=x_\mu\sigma_\mu$ (a $2\times 2$ matrix with spinor
indices) denotes the position and $v(x)$ can be solved explicitly in terms 
of the ADHM data,
\beq
v(x)\!=\!\!\left(\!\!\bea{c}-\ein_n\\u(x)\eea\!\!\right)\phi^{-\hhalf}\!,
\quad u(x)\!=\!(B^\dagger\!-\!x^\dagger)^{-1}\lambda^\dagger,\quad
\phi(x)\!=\!\ein_n\!+\!u^\dagger(x)u(x).
\eeq
The square root $\phi^{\hhalf}(x)$ is well-defined because $\phi(x)$ is a
positive $n\times n$ hermitian matrix. The gauge field is now given by
\beq
A(x)\equiv A_\mu(x)dx_\mu=v^\dagger(x)dv(x)\label{eq:aadhm}
\eeq
and to check its self-duality it is best to use form notation. For the 
field strength two form $F(x)=\half F_{\mu\nu}(x)dx_\mu\wedge 
dx_\nu=dA(x)+A(x)\wedge A(x)$ we find
\beq
F=d(v^\dagger dv)+v^\dagger dv\wedge v^\dagger dv=
dv^\dagger\wedge(1-v\otimes v^\dagger)dv.
\eeq
We note that $v\otimes v^\dagger$ projects to the kernel of $\Delta^\dagger$. 
Assuming that $\Delta$ has no zero eigenvalues such that $\Delta^\dagger\Delta$
is invertible, we find that $1-v\otimes v^\dagger=\Delta(\Delta^\dagger
\Delta)^{-1}\Delta^\dagger$. Indeed, when acting on elements in the kernel 
of $\Delta^\dagger$, left- and right-hand side are equal. Any vector in the 
orthogonal complement of this kernel can be written as $\Delta w$, since 
$<v,\Delta w>=<\Delta^\dagger v,w>=0$, such that left- and right-hand side 
are again equal. Hence
\beq
F=dv^\dagger\Delta\wedge(\Delta^\dagger\Delta)^{-1}\Delta^\dagger dv=
dv^\dagger bdx\wedge(\Delta^\dagger\Delta)^{-1}dx^\dagger b^\dagger dv,
\eeq
where we use the fact that $\Delta^\dagger dv=(d\Delta^\dagger)v=dx^\dagger 
b^\dagger v$, with $b^\dagger=(0~\Eins_2\otimes\Eins_k)$ as a $2k\times 
(n+2k)$ matrix\footnote{The notation $dx^\dagger b^\dagger$ may be a bit 
misleading, it should be read as $(dx^\dagger\otimes\Eins_k)b^\dagger$.}. 
It might seem this does not help that much, but remarkably, using $dx=dx_\mu
\sigma_\mu$ we find $dx\wedge dx^\dagger=\eta_{\mu\nu}dx_\mu\wedge dx_\nu$, 
and since $\eta_{\mu\nu}$ is self-dual, we are done. Not quite so yet! We 
had to take $dx\otimes\Eins_k$ through $(\Delta^\dagger\Delta)^{-1}$ and 
this is only possible if 
\beq
(\Delta^\dagger(x)\Delta(x))^{-1}=\Eins_2\otimes f_x,\label{eq:fdef}
\eeq
where $f_x$ is a hermitian $k\times k$ matrix. This condition, stating that 
$\Delta^\dagger(x)\Delta(x)$ is invertible and commutes with the quaternions, 
is what is known as the quadratic ADHM constraint. It has reduced solving
a set of non-linear partial differential equations to solving a quadratic
matrix equation. 

To construct a charge $k$ caloron with non-trivial holonomy~\cite{FBvB}, we 
place $k$ instantons in the time interval $[0,\beta[$, performing a color 
rotation with $\pl$ for each shift of $t$ over $\beta$, cmp. \refeq{agauge}. 
This is implemented by requiring (suppressing color and spinor indices and 
scaling $\beta$ to 1 in this section)
\beq
\lambda_{pk+k+a}=\pl\lambda_{pk+a},\quad B_{pk+a,qk+b}=B_{pk-k+a,qk-k+b}+
\sigma_0\delta_{pq}\delta_{ab}.
\eeq
Solutions to these equations are parametrized by $\zeta_a$ and $\hat A^{ab}_p$,
\beq
\lambda_{pk+a}=\cP^p_\infty\zeta_a, \quad B_{pk+a,qk+b}=p\sigma_0\delta_{pq}
\delta_{ab}+\hat A^{ab}_{p-q},
\eeq
with $\hat A$ to be determined by the quadratic ADHM constraint. 

\subsection{Fourier-Nahm transformation}\label{sec:nahm}
We introduce the $n$ projectors $P_m$ on the $m$th eigenvalue of $\pl$, 
such that $\pl=\sum_m e^{2\pi i\mu_m}P_m$ and $\lambda_{pk+a}=\sum_m e^{
2\pi ip\mu_m}P_m\zeta_a$. Fourier transformation now leads to
\beq
2\pi i\sum_p e^{2\pi ipz}\hat A^{ab}_{p}=\hat A^{ab}(z),
\quad\sum_p e^{-2\pi ipz}\lambda_{pk+a}=\sum_m\delta(z-\mu_m)P_m\zeta_a.
\eeq
Here $\zeta_a^\dagger$ is again a two-component spinor in the $\bar n$ 
representation of $SU(n)$ and $\hat A^{ab}(z)=\sigma^\mu\hat A^{ab}_\mu(z)$, 
with $\hat A_\mu(z)$ an anti-hermitian $k\times k$ matrix. In terms of 
the latter
\beqa
&&\sum_{p,q} B_{pk+a,qk+b}(x) e^{2\pi i(pz-qz')}=\frac{\delta(z-z')}{2\pi i}
\hat D_x^{ab}(z'),\\ &&\hat D_x^{ab}(z)\equiv\hat D^{ab}(z)-2\pi ix 
\delta^{ab}=\sigma_0\delta^{ab}\ddz+\hat A^{ab}(z)-2\pi ix\delta^{ab},\nonumber
\eeqa
which is the Weyl operator (positive chirality Dirac operator) for the U($k$) 
gauge field $\hat A_\mu(z)\!-\!2\pi ix_\mu\Eins_k$ defined on the circle, 
$z\!\in\![0,1]$, i.e. with periodicity 1 ($\beta^{-1}$ in case $\beta\neq1$). 
The quadratic ADHM constraint now reads
\beq
\Bigl[\sigma_i\otimes\Eins_k,D_x^\dagger(z)D_x(z)+4\pi^2\sum_m\delta(z-\mu_m)
\zeta_a^\dagger P_m\zeta_b\Bigr]=0.
\eeq
Introducing
\beq
2\pi\zeta_a^\dagger P_m\zeta_b\equiv\Eins_2\hat S_m^{ab}-\vec\tau\cdot
\vec\rho_m^{\,ab},\label{eq:rhodef}
\eeq
this leads precisely to the so-called Nahm equation~\cite{Nahm},
\beq
\ddz\hat A_j(z)+[\hat A_0(z),\hat A_j(z)]+\half\veps_{jk\ell}[\hat A_k(z),\hat 
A_\ell(z)]=2\pi i\sum_m\delta(z-\mu_m)\rho_m^{\,j}.\label{eq:nahm}
\eeq
Note that the left-hand side is the difference between the magnetic
and electric field for $\hat A_\mu(z)$, with the right-hand side 
a violation of self-duality.

Although we do not want to go into too much detail here, it is instructive to 
discuss the standard setting of the Nahm transformation~\cite{Nahm,BrvB}. One 
starts from an SU($n$), charge $k$ self-dual gauge field on $\re^4$ with 
periods in all four directions, some of which may be infinite or zero (in the 
latter case effectively leading to a reduced dimension). This extends to a 
family of self-dual U($n$) gauge fields\footnote{One easily checks it does not 
change the field strength $F_{\mu\nu}(x)$.} when adding $-2\pi iz_\mu\Eins_n$ 
to the SU($n$) gauge field $A_\mu(x)$. The index theorem now guarantees there 
is a family of $k$ chiral zero-modes for the Weyl equation $D_z^\dagger\Psi
(x;z)=-\bar\sigma_\mu D_\mu(x;z)\Psi(x;z)=0$, and this allows one to construct 
the gauge field $\hat A^{ab}_\mu(z)=\int d^4x\Psi^a(x;z)^\dagger\partial
\Psi^b(x;z)/\partial z_\mu$. This dual U($k$) gauge field can be shown to 
be self-dual with charge $n$, and is defined again on $\re^4$, but with its 
periods inverted. Remarkably, one can then perform the transformation again, 
and come back to the original gauge field~\cite{BrvB}. 

The ADHM construction performs precisely this last step. For instantons on 
$\re^4$ all periods are infinite and the Nahm transformation reduces 
self-duality to {\em algebraic} equations. Singularities may appear (except 
when all periods are finite), as we have seen from our analysis using Fourier 
transformation of the ADHM data and like in $\re^4$ are related to introducing 
$\lambda$, which encodes the asymptotic behavior of the zero-modes. 

One final result is crucial to appreciate the strength of this formalism, 
\beq
D^\dagger_z D_z=-\Eins_2\otimes D^2_\mu(x;z)-\bar\eta_{\mu\nu}\otimes
[D_\mu(x;z),D_\nu(x;z)],\label{eq:Weit}
\eeq
which uses the fact that $\bar\sigma_\mu\sigma_\nu=\delta_{\mu\nu}\Eins_2+
\bar\eta_{\mu\nu}$. Since $[D_\mu(x;z),D_\nu(x;z)]=F_{\mu\nu}(x)$ and the 
contraction of an anti-self dual tensor ($\bar\eta_{\mu\nu}$) with a self-dual 
tensor ($F_{\mu\nu}$) vanishes, self-duality implies that $D^\dagger_z D_z=
-\Eins_2\otimes D^2_\mu(x;z)$, which therefore commutes with the quarternions. 
This is completely analogous to our discussion for the ADHM construction, and 
proves that the Nahm gauge field $\hat A_\mu(z)$ is self-dual. Doing the Nahm 
transformation for the second time leads us to perform the same calculation
as in \refeq{Weit}, this time for the dual Weyl operator $\hat D_x$, and we 
see that the quadratic ADHM constraint is fully equivalent with stating that 
the dual gauge field is self-dual. The only complication is the possible 
presence of singularities when some of the periods are infinite, in 
particular for the ADHM construction somewhat hiding this profound 
relationship to the Nahm transformation.

\subsection{Some explicit formulae}\label{sec:expl}
One might think all of this is just pushing the same problem around, but as 
we noted before, there is a dramatic simplification due to the dimensional
reduction. In addition, for topological charge 1 the dual gauge field is 
Abelian, further simplifying the Nahm equation. In that case all the 
commutator terms in \refeq{nahm} vanish, and $\hat A_\mu(z)$ is piecewise 
constant, only jumping at the singularities. It is this that has allowed us 
to make progress in finding explicit solutions, together with the ``magic" 
formulae for gauge field, action density, fermion zero-modes and density 
found previously within the ADHM formalism~\cite{Temp,Osb},
\beqa
&&A_\mu(x)=\half\phi^\hhalf(x)\bar\eta_{\mu\nu}^j\partial_\nu\phi_j(x)
\phi^\hhalf(x)+\half[\phi^{-\hhalf}(x),\partial_\mu\phi^\hhalf(x)],
\label{eq:Adef}\\ &&\hskip-5mm\Psi_{iI}^l(x)\!=\!(2\pi)^{-1}\left(
\phi^\hhalf(x)\lambda\partial_\mu f_x\bar\sigma_\mu\veps\right)_{iI}^l,\quad
\Psi^l_{iI}(x)^*\Psi^m_{iI}(x)\!=\!-(2\pi)^{-2}\partial_\mu^2f^{lm}_x,\nonumber
\eeqa
where $l=1,\ldots,k$ labels the zero-modes ($i,I$ are the gauge and spin 
index), 
\beq
\phi(x)=\left(\ein_n-\phi_0\right)^{-1},\quad\phi_\mu\equiv\lambda(\sigma_\mu
\!\otimes\!f_x)\lambda^\dagger,\quad\veps\equiv\sigma_2=i\tau_2.
\eeq
The gauge field for SU(2) further simplifies to $A_\mu(x)=\half\phi(x)\bar
\eta_{\mu\nu}^j\partial_\nu\phi_j(x)$ (which could be viewed as a generalized 
't~Hooft ansatz), since in that case $\phi(x)$ as a $2\times 2$ matrix is 
a multiple of $\Eins_2$. For the action density one finds~\cite{Osb},
\beq
\Tr F_{\mu\nu}^2(x)=-\partial^2_\mu\partial^2_\nu\log\det f_x.
\eeq
It is thus very convenient to first find the matrix $f_x$, as defined in 
\refeq{fdef}. For calorons, after Fourier transformation, it is replaced by 
the Green's function $\hat f_x(z,z')$, which satisfies the equation~\cite{FBvB}
\beq
\left\{\left(\frac{\hat D_\mu(z;x)}{2\pi i}\right)^{\!\!2}\!+
\frac{1}{2\pi}\sum_m\delta(z\!-\!\mu_m)\hat S_m\right\}\!\hat f_x(z,z')=
\ein_k\delta(z\!-\!z').\label{eq:df}
\eeq
The gauge field, as determined through $\phi_\mu(x)$, see \refeq{Adef}, 
is read off from
\beq
\phi_\mu(x)=\sum_{m,m'}P_m\zeta_a\sigma_\mu\hat f_x^{ab}(\mu_m,\mu_{m'})
\zeta^{\dagger}_bP_{m'},
\eeq
whereas the fermion zero-modes satisfying a generalized boundary condition
$\hat\Psi_z^a(t+1,\vec x)=e^{2\pi iz}\pl\hat\Psi_z^a(t,\vec x)$, cmp.
\refeq{agauge}, are given by~\cite{ZM2,BNvB}
\beq
\hat\Psi_z^a(x)=(2\pi)^{-1}\phi^\hhalf(x)\sum_m P_m\zeta_b\bar\sigma_\mu
\veps\partial_\mu\hat f^{ba}_x(\mu_m,z).\label{eq:zmdef}
\eeq
For $z=\half$ this gives the usual anti-periodic boundary conditions for
fermions at finite temperature, but the general $z$ dependence will turn out 
to be extremely useful as a diagnostic tool. Finally, the zero-mode density 
reads
\beq
\hat\Psi_z^a(x)^\dagger\hat\Psi_z^b(x)=-(2\pi)^{-2}
\partial_\mu^2\hat f^{ab}_x(z,z).\label{eq:zmdens}
\eeq

To compute $\hat f_x(z,z')$ we note that we can always transform to a gauge 
where $\hat A_0(z)\equiv 2\pi i\xi_0$ is constant, after which $\hat g(z)
\equiv\exp(2\pi i(\xi_0-x_0\Eins_k)z)$ transforms $\hat A_0(z)-2\pi ix_0$ to 
zero. This turns \refeq{df} into
\beq
\left\{-\frac{d^2}{dz^2}+V(z;\vec x)\right\}
\!f_x(z,z')=4\pi^2 \ein_k\delta(z\!-\!z'),\label{eq:dftrans}
\eeq
with $f_x(z,z')$ and $V(z;\vec x)$ given by
\beqa
&&\hskip-4mm f_x(z,z')\!\equiv\!\hat g(z)\hat f_x(z,z')\hat g^\dagger(z'),
\quad V(z;\vec x)\!\equiv\!4\pi^2\vec R^2(z;\vec x)\!+\!2\pi\!\!\sum_m
\delta(z\!-\!\mu_m)S_m,\nonumber\\&&\hskip-4mm R_j(z;\vec x)\equiv x_j
\ein_k-(2\pi i)^{-1}\hat g(z)\hat A_j(z)\hat g^\dagger(z),\quad S_m
\equiv\hat g(\mu_m)\hat S_m\hat g^\dagger(\mu_m).\label{eq:Rdef}
\eeqa
Periodicity is now only up to the gauge transformation $\hat g(1)$. 
In particular when $\vec R^{\,2}(z;\vec x)$ is piecewise constant 
explicitly computing the Green's function becomes doable. Nevertheless,
in general terms one finds
\beqa
&&\hskip-4mmf_x(z,z')=4\pi^2\left\{W(z,z_0)\left(\theta(z'-z)\ein_{2k}-
(\ein_{2k}-\cF_{z_0})^{-1}\right)W^{-1}(z',z_0)\right\}_{12},\nonumber\\
&&\hskip-4mmW(z_2,z_1)\equiv\Pexp\int_{z_1}^{z_2}\pmatrix{0&\ein_k\cr
V(z;\vec x)&0\cr}dz,\quad\cF_{z_0}\equiv\hat g^\dagger(1)W(z_0\!+\!1,z_0)
.\label{eq:Wdef}
\eeqa
where the $(1,2)$ component on the right-hand side for $f_x(z,z')$ is with
respect to the $2\times 2$ block matrix structure. This equation for 
$f_x(z,z')$ is valid for $z'\in[z,z+1]$, but can be extended with the 
appropriate periodicity. We can now also find an explicit result for the 
action density
\beq
\quad\Tr F_{\mu\nu}^2(x)=\partial_\mu^2\partial_\nu^2\log\psi(x),\quad
\psi(x)\equiv\det\left(ie^{-\pi ix_0}(\Eins_{2k}-\cF_{z_0})/\sqrt{2}\right).
\label{eq:Sdens}
\eeq
Note that $z_0$ can be chosen at will, e.g. for $z_0=\mu_m\!+0$ we find
\beqa
&&\cF_{\mu_m}=T_mH_{m-1}\cdots T_2H_1T_1\hat g^\dagger(1)H_n
T_nH_{n-1}\cdots T_{m+1}H_m,\label{eq:THs}\\
&&T_m\equiv\pmatrix{\ein_k&0\cr 2\pi S_m&\ein_k\cr},\quad 
H_m\equiv\Pexp\int_{\mu_m}^{\mu_{m+1}}\pmatrix{0&\ein_k\cr 
4\pi^2\vec R^2(z;\vec x)&0\cr}dz.\label{eq:Hm}\nonumber
\eeqa
The main pay off to have these expressions in terms of possibly unknown
solutions to the Nahm equations, is that it allows us to look at the 
far field limit. When constituent monopoles are well separated, the 
charged components of the field in the core decay exponentially, being 
left with the Abelian fields in the far field region. This is related 
to the high temperature limit, in which case the cores collapse to zero 
size.

\subsection{Limits and special cases}\label{sec:spec}
To extract exponential factors it turned out to be convenient to define 
\beq
f_m^\pm(z)\!=\Pexp\left[\pm 2\pi\!\int_{\mu_m}^z\!\!\!\!R_m^\pm(z)dz\right],
\quad R_m^\pm(z)^2\pm\frac{1}{2\pi}\ddz\,R_m^\pm(z)\!=\!\vec R^{\,2}(z;\vec x).
\label{eq:Riccati} 
\eeq
Since $\vec R(z;\vec x)\to\vec x\ein_k$ for $|\vec x|\to\infty$, we find in 
this limit that $R_m^\pm(z)\to|\vec x|\ein_k$ and $f_m^\pm(z)\to\exp\left(
\pm 2\pi|\vec x|(z-\mu_m)\ein_k\right)$. For $z,z'\in[\mu_m,\mu_{m+1}]$ we 
can now write $W(z,z')=W_m(z)W_m^{-1}(z')$, see \refeq{Wdef}, with
\beq
W_m(z)\equiv\pmatrix{f_m^+(z)&f_m^-(z)\cr 
2\pi R^+_m(z)f_m^+(z)&-2\pi R^-_m(z)f_m^-(z)\cr}.
\eeq
This has allowed us to show that (ff stands for far field limit)
\beq
f^\ff_x(\mu_m,\mu_m)=2\pi\Sigma_m^{-1},~\mbox{and}~
f^\ff_x(z,z)=\pi R_m^{-1}(z)\mbox{~for~}\mu_m\!<\!z\!<\!\mu_{m+1}
, \label{eq:fflim}
\eeq
where $\Sigma_m\equiv R_m^-(\mu_m)+R_{m-1}^+(\mu_m)+S_m$ and $R_m(z)\equiv
\half(R_m^+(z)+R_m^-(z))$. For this to hold $\vec x$ has to be far removed 
from any constituent. In addition $f^\ff(\mu_m,\mu_{m+1})=0$, which implies 
that for SU(2) and SU(3) only the Abelian components of the gauge field 
survive in this limit. 

Useful is also the so-called zero-mode limit (zm), which assumes $\vec x$ to 
be far removed from any constituent monopole other than of type $m$, 
distinguished by their magnetic charge and mass $8\pi^2\nu_m$, with 
$\nu_m\equiv\mu_{m+1}-\mu_m$, see the next section. In this limit one finds 
up to {\em exponential} corrections~\cite{BNvB} for $\mu_m\leq z'\leq z\leq
\mu_{m+1}$ ($f_x(z',z)=f_x^\dagger(z,z')$ for $z'>z$) 
\beqa
&&\hskip-4mm f^\zm_x(z,z')=\pi\left(f_m^-(z)f_m^-(\mu_{m+1})^{-1}\!-f_m^+(z)
f_m^+(\mu_{m+1})^{-1}Z_{m+1}^-\right)\times\label{eq:zmlim}\\
&&\hskip-4mm \left(f_m^-(\mu_{m+1})^{-1}\!-\!Z_m^+f_m^+(\mu_{m+1})^{-1}Z_{m+1}^-
\right)^{-1}\!\!\left(f_m^-(z')^{-1}\!-\!Z_m^+f_m^+(z')^{-1}\right)R_m^{-1}(z'),
\nonumber
\eeqa
with $Z_m^-\equiv\ein_k-2\Sigma_m^{-1}R_{m-1}(\mu_m)$ and $Z_m^+\equiv\ein_k-2
\Sigma_m^{-1}R_m(\mu_m)$. One concludes that the zero-modes (see \refeq{zmdef} 
and \refeq{zmdens}) are localized exponentially to the $m$th constituent 
monopoles (for $z$ away from the interval boundaries).

\begin{figure}[htb] 
\vskip3mm\hskip6.4cm $y$ \vskip1.3cm \hskip9.5cm $\xi$ \vskip1.7cm
\includegraphics{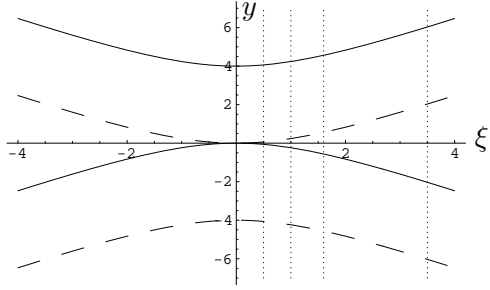}
\vskip2mm
\caption{Constituent locations $y$ based on \refeq{Ym} as a function of 
$\xi=\xi_1=-\xi_2$ for $y_2= -y_1=\nu_1=\nu_2=\half$ and $\rho_1=\rho_2=2$. 
Dashed versus full lines distinguish the magnetic charge of the constituents. 
The dotted lines apply to Figs.~\ref{fig:3}.}\label{fig:4}
\end{figure} 

Explicit solutions are found when $\vec R(z;\vec x)\equiv\vec x-\vec Y(z)$ is 
piecewise constant, i.e. $\vec Y(z)\equiv\vec Y_m$ for $z\in[\mu_m,\mu_{m+1}]$,
in which case $|\vec R|=R_m^\pm=R_m$. Apart from $k=1$, this is so for a class 
of axially symmetric multi-caloron solutions~\cite{FBvB}, in which case $3k$ 
eigenvalues of $\vec Y_m$ give the locations of the type $m$ constituent 
monopoles. The expression for the zero-mode in that case can be used to show 
that in the high temperature limit the zero-mode densities reduce to delta 
functions localized at these constituent locations, thus establishing that 
the constituents in this limit are point-like monopoles. Particularly simple 
axially symmetric solutions are found for SU(2),
\beq
\zeta^a_{iI}=\rho_a\delta_{iI},\quad\vec Y^{ab}_m=\vec e(\delta_{ab}\xi_a+
y_m\rho_a\rho_b),\quad\nu_1 y_1+\nu_2 y_2=0,\quad\rho_a>0.\label{eq:Ym}
\eeq
For charge 2, choosing $\rho_1=\rho_2=2$ and equal mass constituents 
($\nu_1=\nu_2=\half$) the constituent locations $y$ are plotted as a 
function of $\xi=\xi_1=-\xi_2$ in Fig.~\ref{fig:4}.
We note that opposite charges are found to alternate. The properties of these
solutions will be further discussed in the next section. 

More generally, for SU(2) and charge 2 we may borrow from the study of
monopoles~\cite{Nahm,Pana} the general solution of the Nahm equations 
for $z\!\in\![\mu_1,\mu_2]$
\beq
\hat A_j(z)\equiv 2\pi i\hat g^\dagger(z)h^\dagger\Bigl(a_j\ein_2+\cD\cR_{jb}
f_{b}(4\pi\cD z)\tau_b\Bigr)h\hat g(z),\label{eq:fnahm}
\eeq
where $f_j(z)$ are the Jacobi elliptic functions
\beq
f_1(z)=\frac{\k'}{cn_\k(z)},\quad f_2(z)=\frac{\k'sn_\k(z)}{cn_\k(z)},
\quad f_3(z)=\frac{dn_\k(z)}{cn_\k(z)}.
\eeq
Here $\vec a$ is the center of mass for monopoles of a given type, $\cR$ 
and $h$ are arbitrary spatial and gauge rotations, $\cD$ is a scale 
parameter, and $0\leq\k\leq1$ ($\k'\equiv\sqrt{1-\k^2}$) playing the role
of a shape parameter, as will be discussed in the next section. In general 
all these parameters differ on the second interval where in addition $z$ is 
shifted to $z-\half$, but they are to be related through the discontinuities 
in the Nahm equation, \refeq{nahm}. This tends to be rather restrictive, and 
is the point where the construction for calorons deviates from that for 
static monopoles.

Quite remarkably, although $\vec R(z;\vec x)$ is no longer piecewise constant,
the function $\Tr R^{-1}_m(z)$ {\em is independent} of $z\in[\mu_m,\mu_{m+1}]$. 
This is a highly non-trivial consequence of the Nahm equations (from the point 
of view of integrability, it gives a constant of motion). The physical 
significance here is that it is directly related to the zero-mode density 
(summed over the zero-modes) in the high temperature limit, see 
\refeq{zmdens} and \refeq{fflim},
\beq
\Bigl(\sum_a\hat\Psi_z^a(x)^\dagger\hat\Psi_z^a(x)\Bigr)^\ff=
-\partial_i^2\cV_m(\vec x),\quad\cV_m(\vec x)\equiv(4\pi)^{-1}\Tr R^{-1}_m(z).
\label{eq:Vmdef}
\eeq
Miraculously we have been able to calculate $\cV_m(\vec x)$ 
exactly~\cite{BNvB}, from which we will be able to draw conclusions 
on the pointlike nature of the constituents.

\section{Properties of solutions}\label{sec:prop}
\begin{figure}[htb]
\vspace{3.3cm} 
\includegraphics{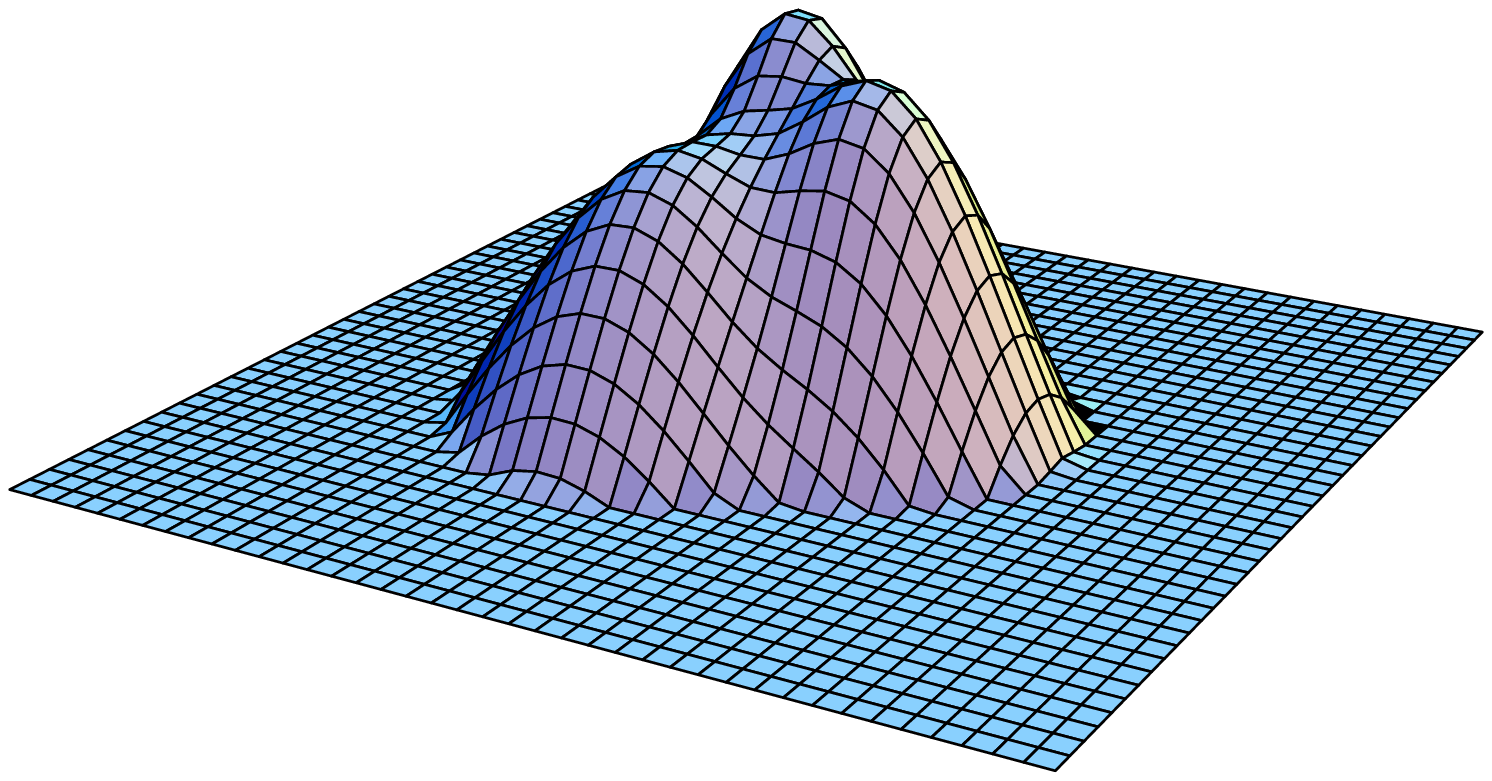}
\includegraphics{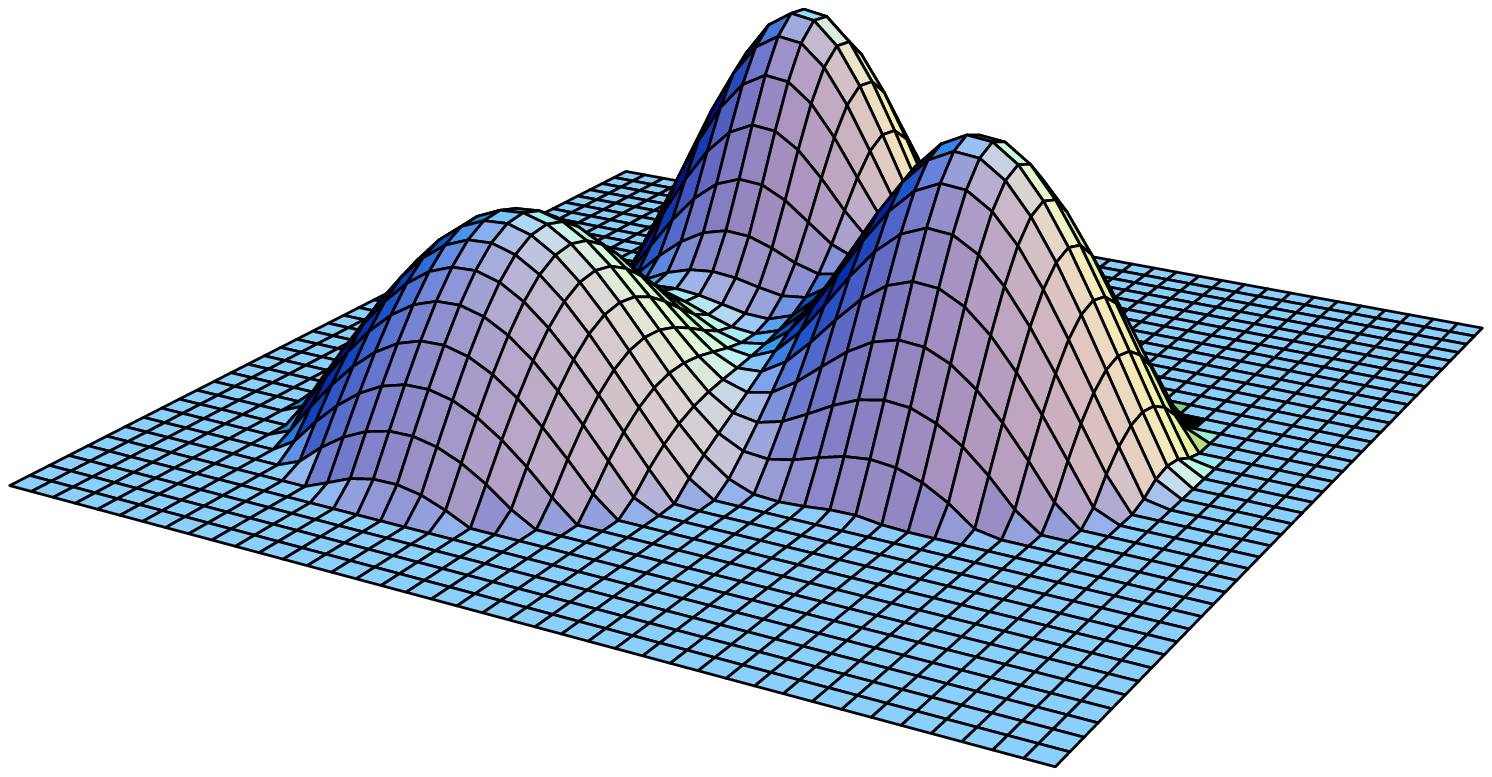}
\includegraphics{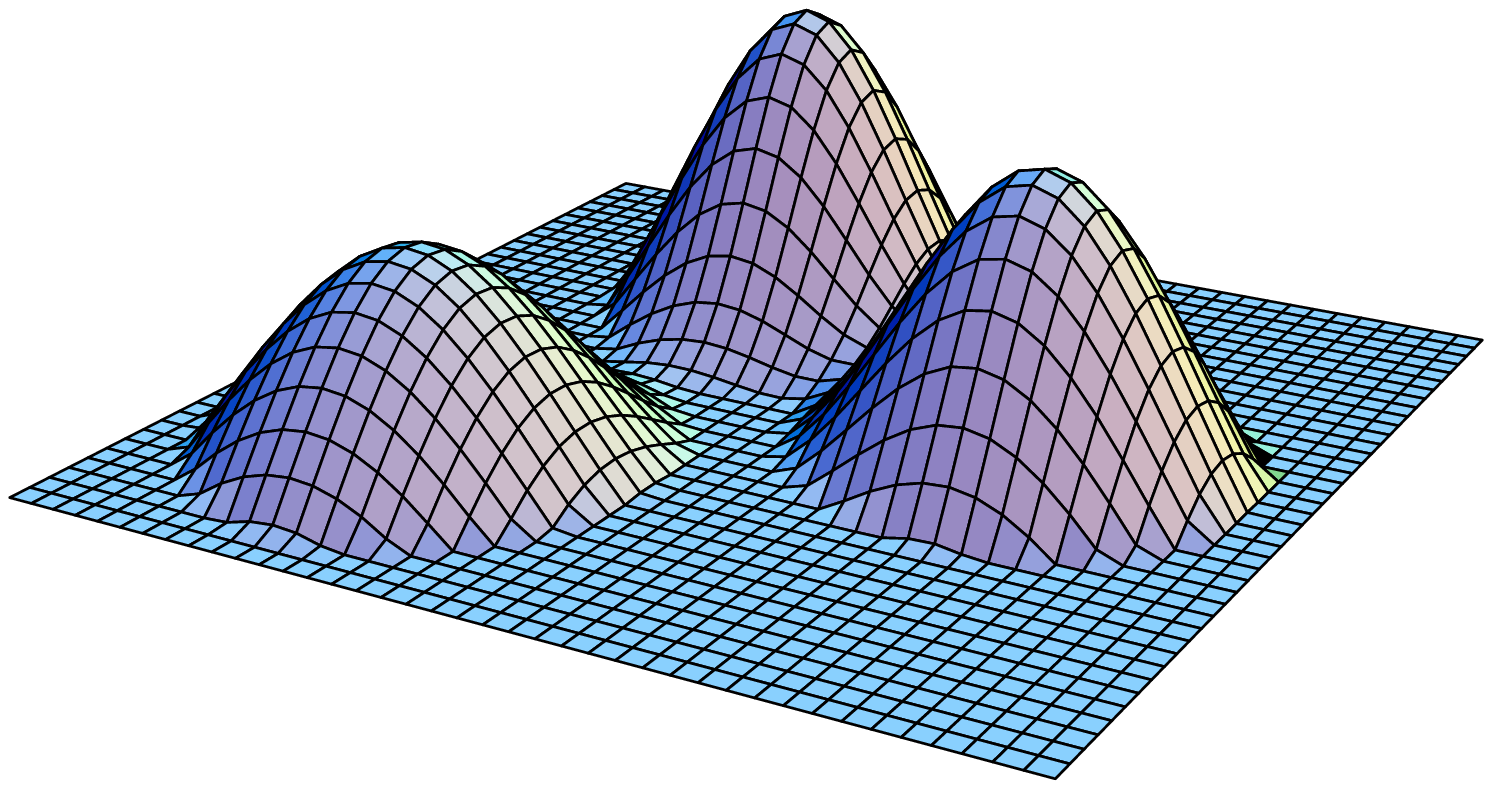}
\caption{Action densities for the $SU(3)$ caloron with $(\mu_1,\mu_2,\mu_3)=
(-17,-2,19)/60$ and $(\nu_1,\nu_2,\nu_3)=(0.25,0.35,0.4)$ at $t=0$ in the 
plane of the three constituents for $1/\beta=1.5$, 3 and 4 (from left to 
right) on equal logarithmic scales.}\label{fig:5}
\end{figure}

We start our tour of SU($n$) caloron solutions with those of charge 1.
In this case the action density has a particularly simple form~\cite{KvBn} 
\beqa
&&-\half\Tr F_{\mu\nu}^{\,2}(x)=-\half\partial_\mu^2\partial_\nu^2
\log\psi(x),\quad\psi(x)=\half\tr(\cA_n\cdots \cA_1)-\cos(2\pi t/\beta),
\nonumber\\ &&\cA_m\equiv\frac{1}{r_m}\left(\!\!\!\bea{cc}r_m\!\!&|\vec
\rho_{m+1}|\\0\!\!&r_{m+1}\eea\!\!\!\right)\left(\!\!\!\bea{cc}\cosh(2\pi
\nu_m r_m/\beta)\!\!&\sinh(2\pi\nu_m r_m/\beta)\\ \sinh(2\pi\nu_m r_m/\beta)
\!\!&\cosh(2\pi\nu_m r_m/\beta)\eea\!\!\!\right).\label{eq:exactS}
\eeqa
Here $\nu_m=\mu_{m+1}-\mu_m$, $r_m=|\vec r_m|$, $\vec r_m=\vec x-\vec y_m$ 
and $\vec\rho_m=\vec y_m-\vec y_{m-1}$, where $\vec y_m$ ($\vec y_{n+m}\equiv
\vec y_m$) are the constituent locations\footnote{This can be derived from
Eqs.~(\ref{eq:Sdens},\ref{eq:THs}) using $S_m=\hat S_m=|\vec\rho_m|$ (see
\refeq{rhodef}) and $\cA_m=B_{m+1}T_{m+1}H_mB_m^{-1}$, where $B_m^{11}=
B_m^{22}=0$, $B_m^{12}=1$ and $B_m^{21}=4\pi r_m$.}. A few typical 
examples for SU(2), where $|\vec\rho_1|=\pi\rho^2/\beta$ in terms of 
the instanton scale parameter $\rho$, are shown in Fig.~\ref{fig:2} and for 
SU(3) in Fig.~\ref{fig:5} (we apologize for a somewhat awkward choice of 
conventions in naming the vectors $\vec\rho_m$ and scale parameters $\rho_a
=|\zeta_a|$). From this it is already clear that there are $n$ lumps, 
centered at $\vec y_m$ and that when well-separated they are static 
and spherically symmetric. The self-duality then guarantees each lump is a 
basic BPS monopole, which can be seen to contribute $8\pi^2\nu_m$ to the 
action, correctly summing to a total action $8\pi^2$. Interesting properties
on the geometry of the moduli space and alternative approaches can be found 
in Refs.~\cite{LeLu,PLB,Kra}. 

It is instructive to provide further evidence for the monopole nature of 
these constituent lumps. For this we take the far field limit (ff) of 
\refeq{exactS}. In this limit $\vec x$ is assumed to be far from all 
constituent locations $\vec y_m$ and we find 
\beq
\psi^\ff(x)=\half\prod_{m=1}^n\frac{\left(r_m+r_{m+1}+|\vec\rho_{m+1}|
\right)}{2r_m}e^{2\pi\nu_m r_m/\beta}.\label{eq:abelS}
\eeq
Note that there is no longer any time dependence and the far field limit is 
equivalent to the high temperature limit $\beta\to0$, where the monopole mass 
becomes infinite and the non-Abelian core collapses to zero size (the range of 
the exponentially decaying charged fields shrinks to zero). As we will argue,
using $A_\mu^{\rm D}(\vec x)$ as the gauge field for the basic self-dual 
Abelian Dirac monopole, one finds in a suitable gauge for the Abelian far field 
embedded in SU($n$), $A^{ab}_\mu(\vec x)=\delta_{ab}\tilde A_\mu^a(\vec x)$, 
with $\tilde A_\mu^m(\vec x)=\half A_\mu^{\rm D}(\vec r_m)-\half A_\mu^{\rm D}
(\vec r_{m-1})$, where $\vec E^{\rm D}(\vec x)=\vec B^{\rm D}(\vec x)=\vec x
|\vec x|^{-3}$ are electric and magnetic fields of this self-dual Dirac 
monopole. One 
verifies\footnote{It is best to first check $-\half\partial_i^2\partial_j^2
\log[r_1^{-1}r_2^{-1}(r_1+r_2+|\vec\rho_2|)^2]=\left(r_1^{-3}\vec r_1-
r_2^{-3}\vec r_2\right)^2$, relevant for SU(2). The exponential terms in 
\refeq{abelS} do not contribute, except for $\sum_m\delta(\vec r_m)8\pi^2
\nu_m/\beta$. However, in the core \refeq{abelS} should not be used since the 
singularity in $\psi^\ff(x)$ is simply due to approximations made outside 
the core.} that indeed $-\half\Tr F_{\mu\nu}^{\,2}(x)=-\half\partial_i^2
\partial_j^2\log\psi^\ff(x)$ when substituting this gauge field.

To give a little more insight in why the gauge field takes the above form, 
we remark that another way to define the location of an SU(2) monopole is to 
find the zeros of the Higgs field, or for SU($n$) to find where two of its 
eigenvalues coincide. At these points the broken gauge symmetry is partially
restored to include an unbroken SU(2) subgroup. In the case of the caloron we 
of course need to find coinciding eigenvalues of the Polyakov loop\footnote{In 
technical terms the locations of constituents are found where $\log P(\vec x)$ 
takes values on the boundary of the Weyl chamber~\cite{Berl}.}, $P(\vec x)$
\cite{Dubna}. The gauge field $\tilde A^m(\vec x)$ defined above is 
associated to the Abelian component of the embedded SU(2) monopole associated 
to the unbroken subgroup. The Polyakov loop therefore is a useful diagnostic 
tool particularly when monopoles are too close to be seen as separate lumps. 
For SU(2) the constituents are simply found where $P(\vec x)=\pm\Eins_2$ 
\cite{CCal}.

\begin{figure}[htb]
\vspace{3.1cm} 
\includegraphics{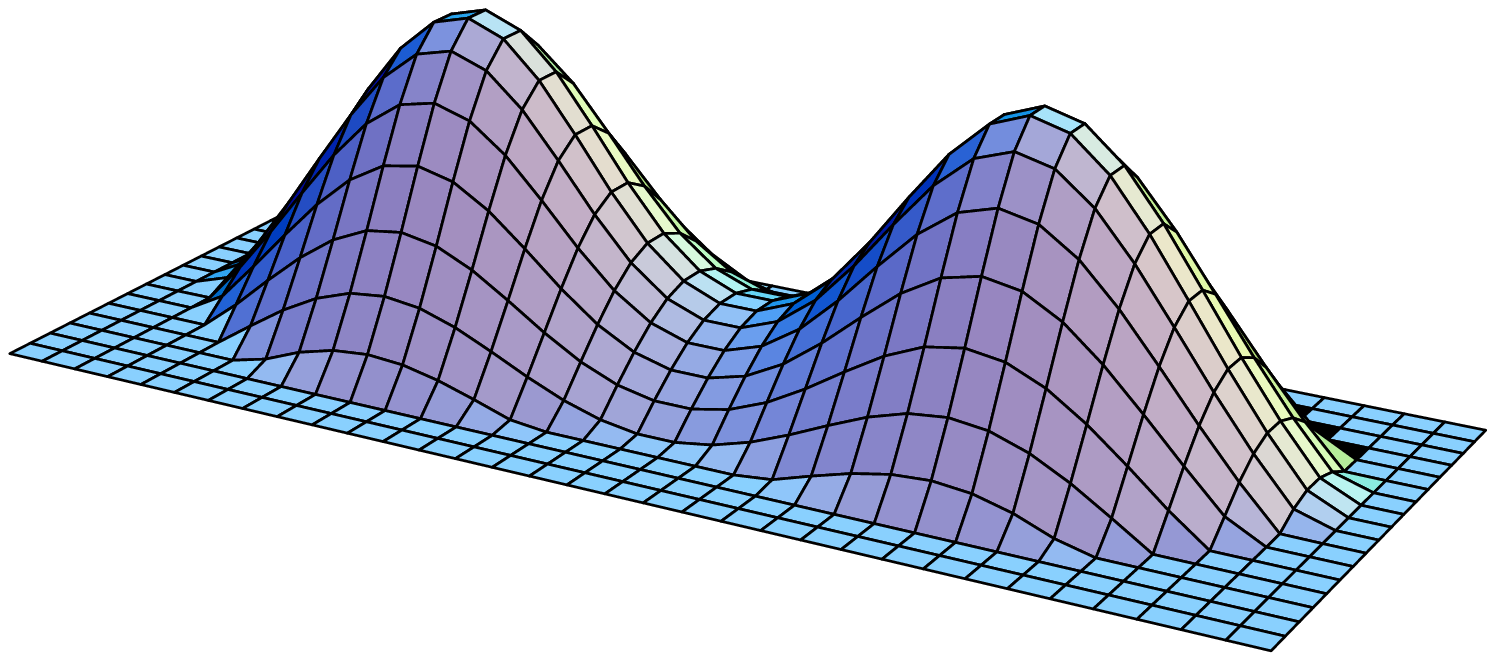}
\includegraphics{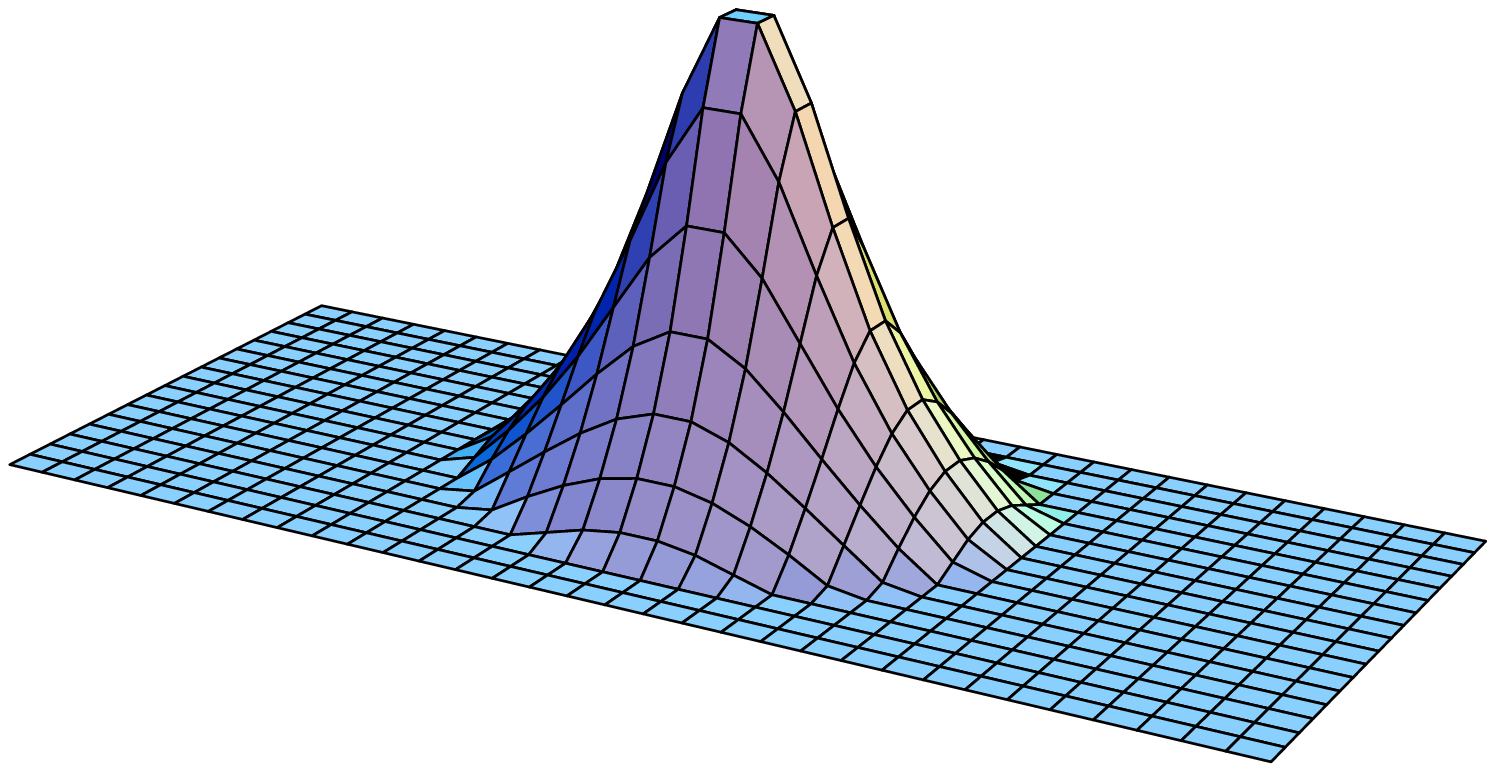}
\includegraphics{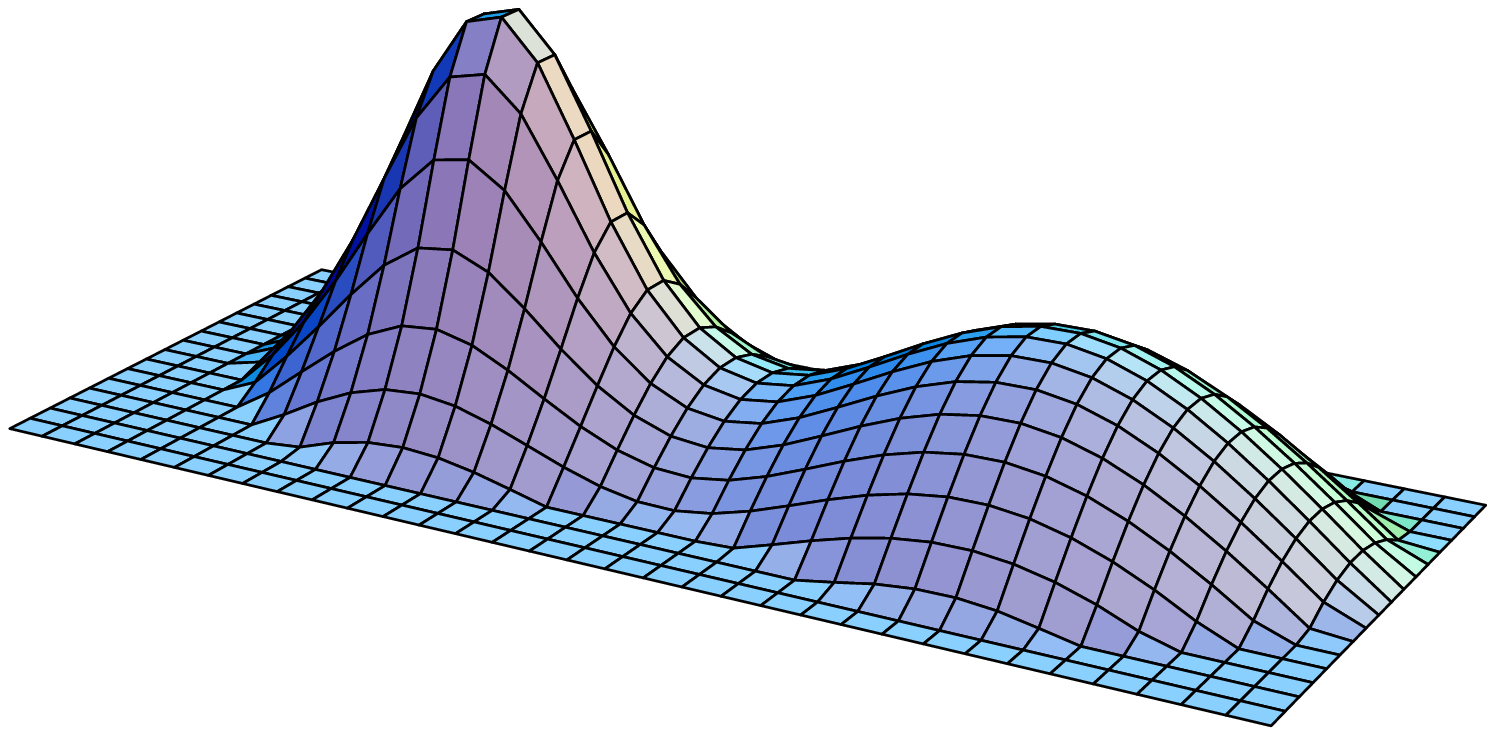}
\caption{The action density for SU(2) calorons with $\rho=\beta$ and (from 
left to right) $\mu_2=0$ (i.e. trivial holonomy), 0.125 and 0.25 (i.e. $\Tr
\pl=0$), at $t=0$ in a plane through the constituent locations.}\label{fig:6}
\end{figure}

Important is that when coinciding eigenvalues occur at infinity, that is in 
$\pl$, some constituent masses will vanish. For trivial holonomy only the 
$n$th constituent is massive and the action density shows a single lump, the 
usual (deformed) instanton, cmp. Fig.~\ref{fig:6}. In the presence of 
massless constituents one cannot take the far field limit, due to the infinite 
range of the fields in these massless cores. One may, however, move massless 
constituents off to infinity. For SU(2) this requires one to take $\rho\to
\infty$, and one is left with a BPS monopole in a singular gauge to adjust 
the mismatch in boundary conditions between caloron and monopole, or put 
differently, to compensate for the magnetic charge that is pushed past 
infinity~\cite{Ros}. 

As we have explained in the introduction, chiral zero-modes play an extremely 
important role in the physics of instantons. They will also turn out to be 
very useful as a diagnostic tool for studying the properties of the caloron
solutions. The Atiyah-Index theorem has taught us there is one chiral zero-mode
for a charge 1 caloron. But when the caloron has ``dissociated" in $n$ 
constituents the natural question is if the zero-mode is going to follow this.
There is a compelling reason to expect it will opt for being localized on one 
constituent only. When well separated, the constituents become BPS monopoles. 
These are known to have zero-modes, but only for a given range of $z$ values
in terms of the Higgs expectation value, as dictated by the Callias index 
theorem~\cite{Call}. We will show how this determines to which constituent 
the zero-mode will be localized. Even more useful is that we can change this 
by introducing an arbitrary phase in the boundary conditions for the fermions, 
which reads in the algebraic gauge (\refeq{agauge})
\beq
\hat\Psi_z(t+\beta,\vec x)=\exp(2\pi iz\beta)\pl\hat\Psi_z(t,\vec x).
\label{eq:cycle}
\eeq
Physical fermions at finite temperature are required to be anti-periodic,
$\Psi(t,\vec x)=\hat\Psi_{z=1/2\beta}(t,\vec x)$. That $z$ determines 
the location of the zero-mode is seen as follows. The gauge transformation 
$g(t)=\exp(2\pi izt)\exp(t A^\infty_0)$ makes the fermions periodic 
at the expense of changing $A_0=0$ at spatial infinity to $A_0=A^\infty_0-
2\pi iz$, cmp. \refeq{pol}. This acts as an effective mass term in the Dirac 
equation, which differs for each of the gauge components. 
\begin{figure}[htb]
\vskip2.7cm
\includegraphics{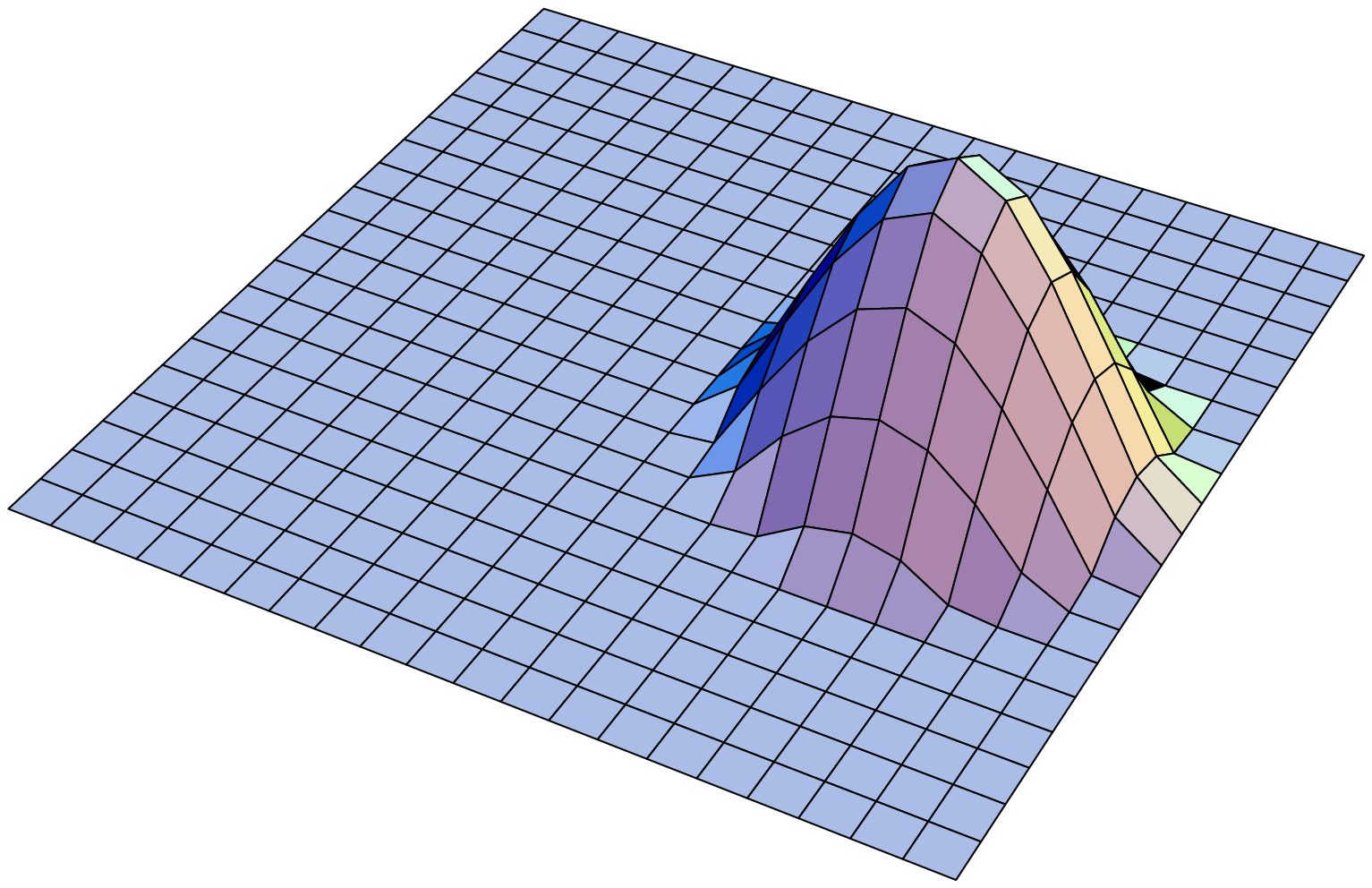}
\includegraphics{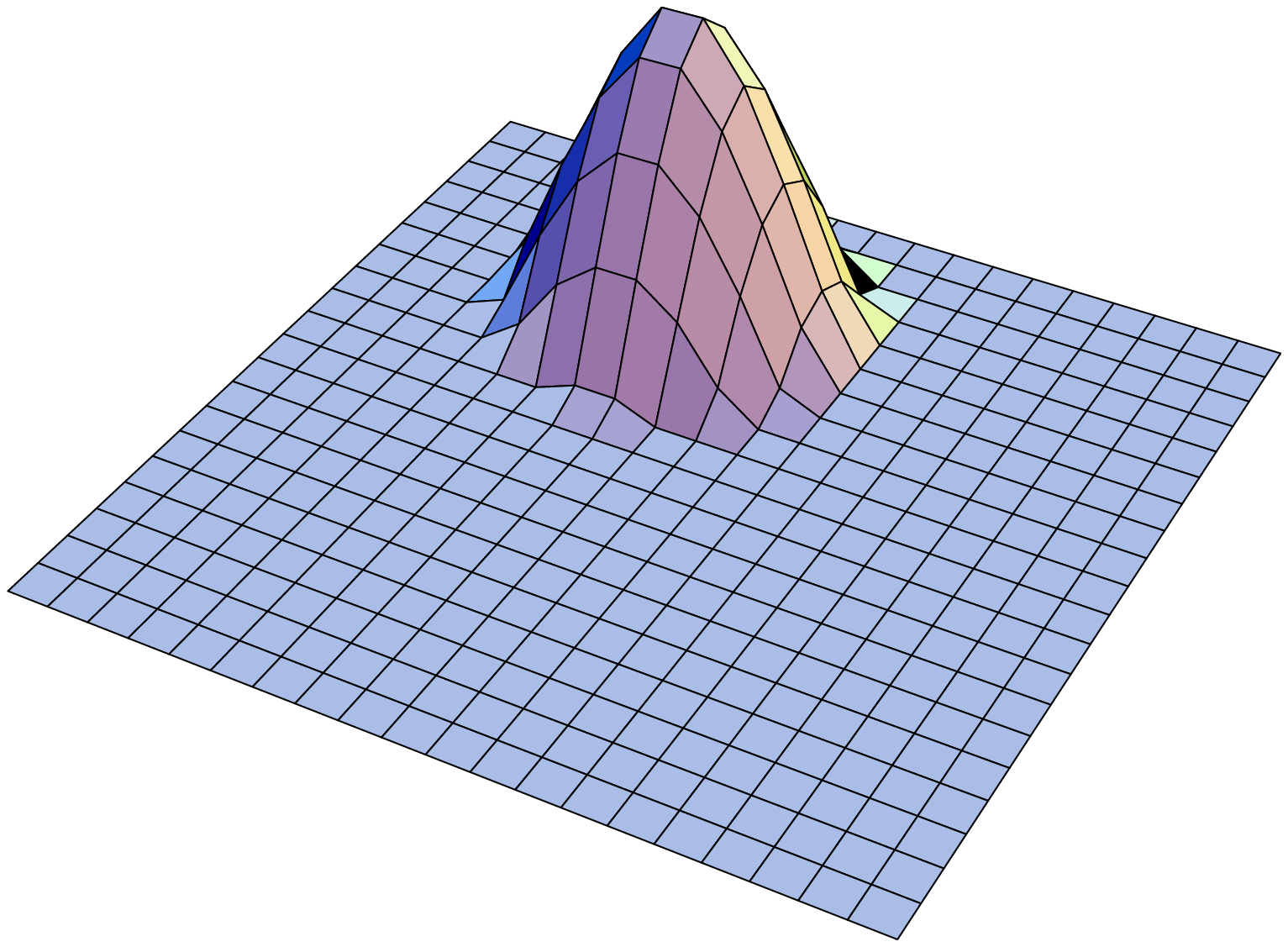}
\includegraphics{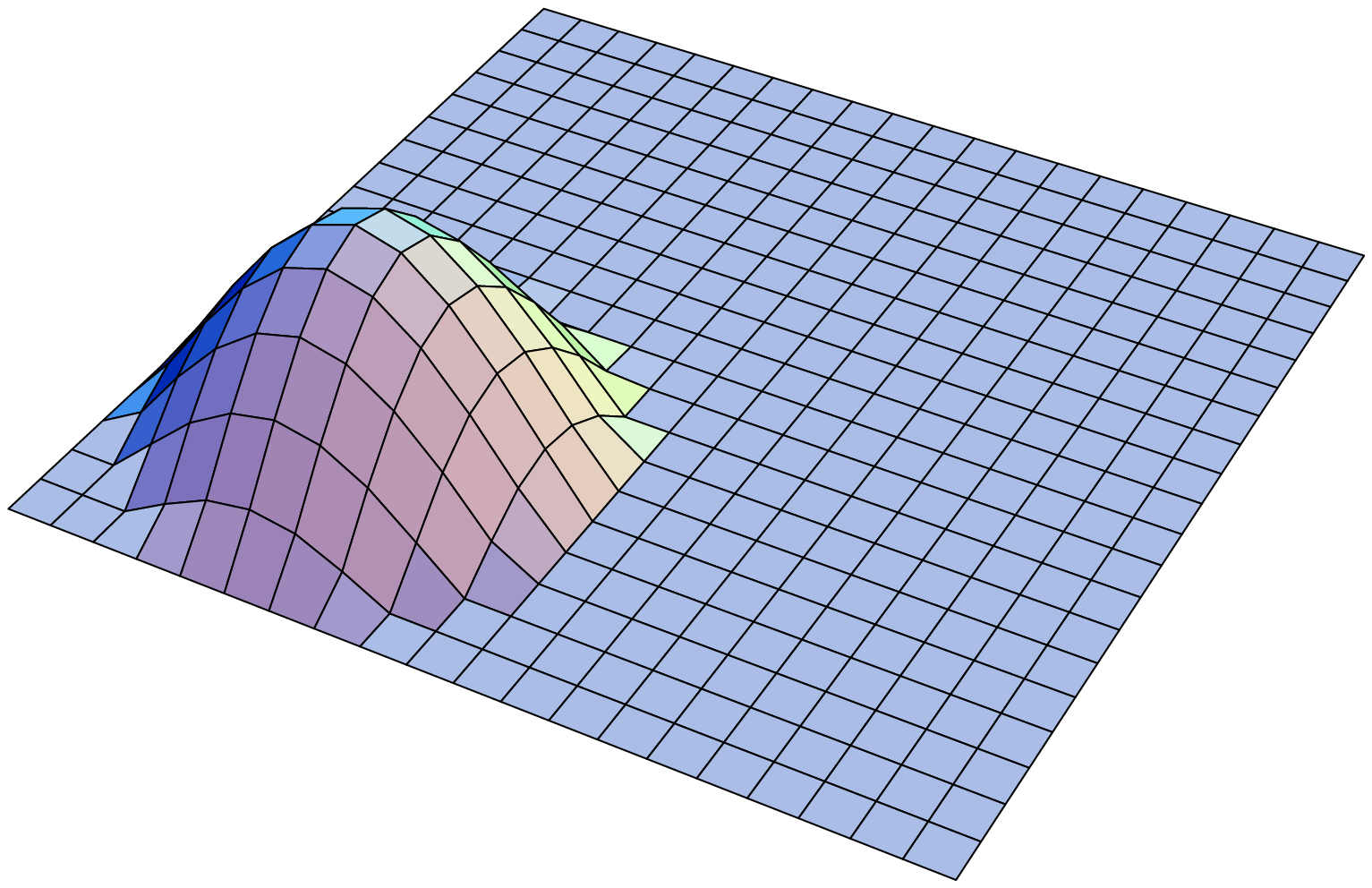}
\hskip0.1cm$z=12/60$\hskip2.6cm$z=30/60$\hskip2.6cm$z=48/60$
\vskip2.7cm
\includegraphics{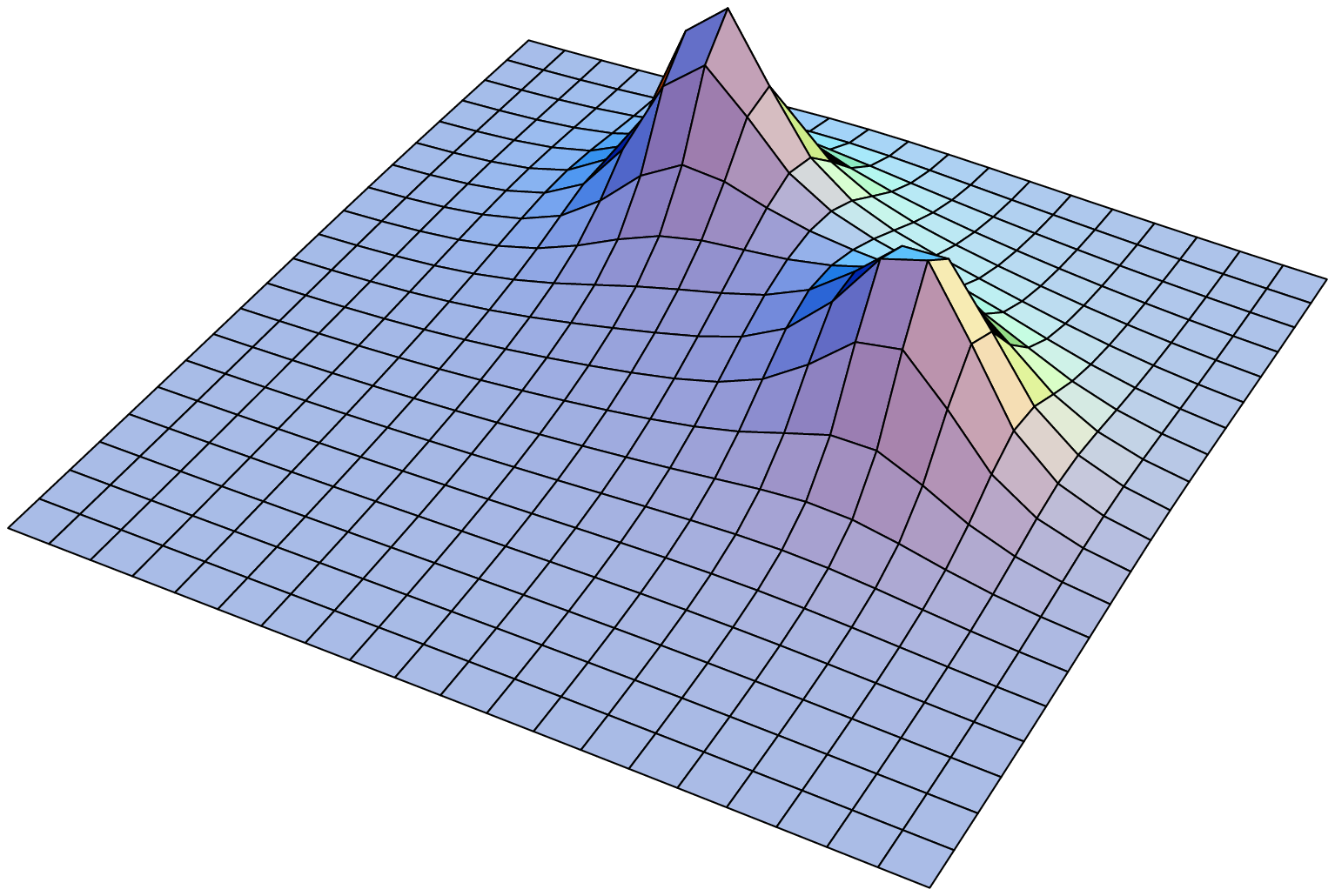}
\includegraphics{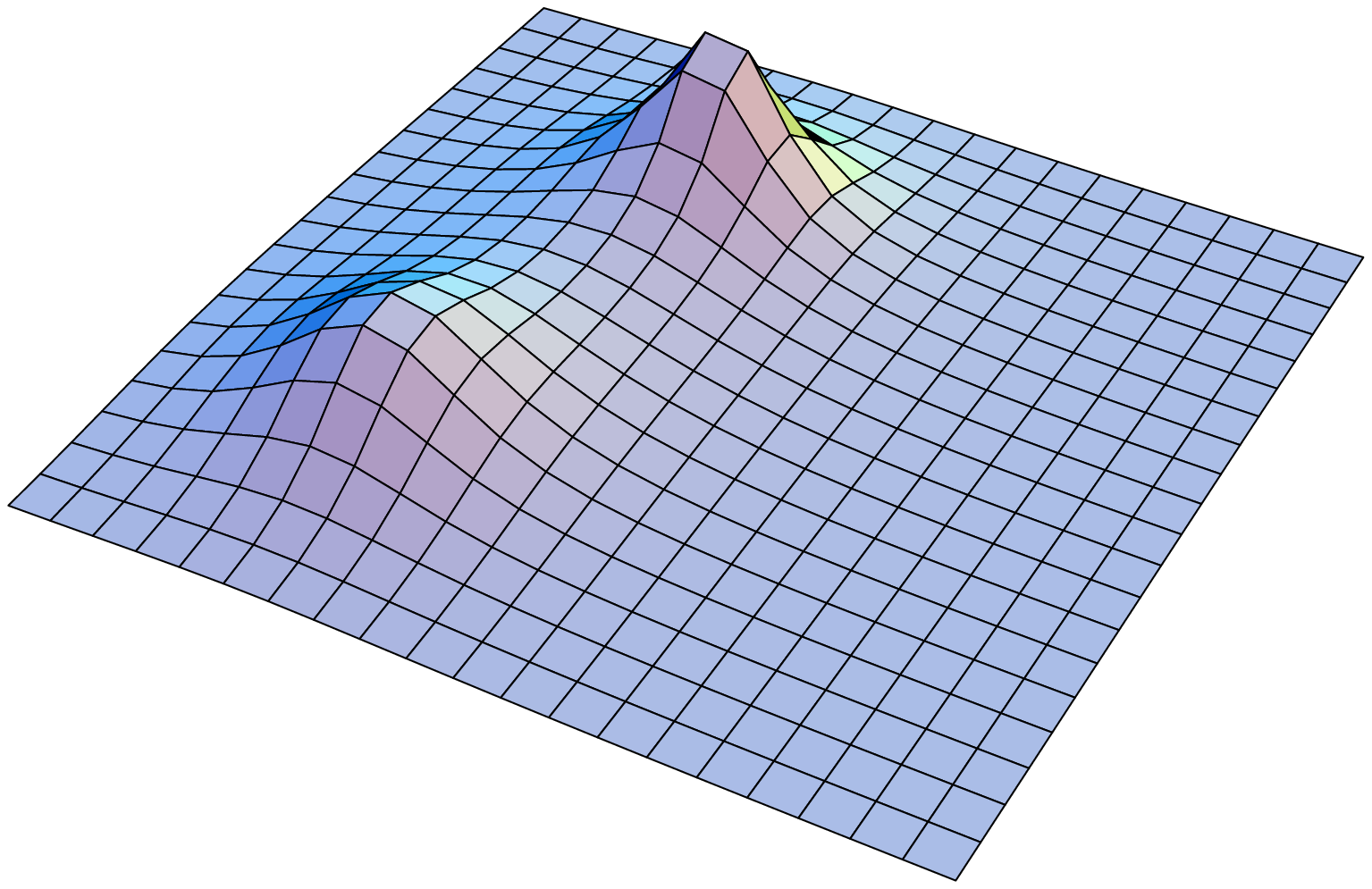}
\includegraphics{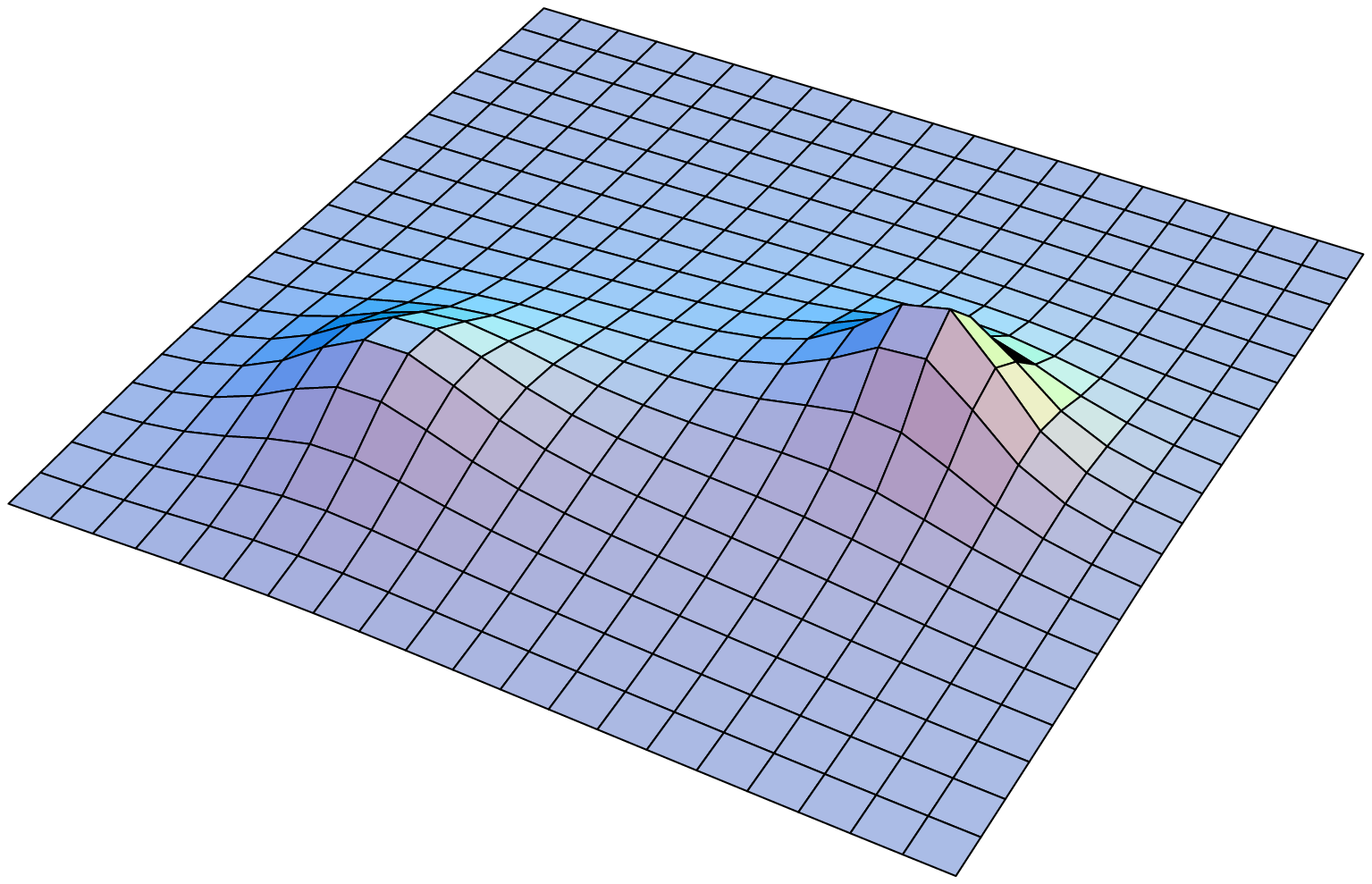}
\hskip0.1cm$z=19/60$\hskip2.6cm$z=43/60$\hskip2.6cm$z=58/60$
\caption{Normalized zero-mode densities at $\beta=1$, for the $SU(3)$ caloron
of Fig.~\ref{fig:5} shown for $z=\mu_j$ on equal linear scales (bottom) and on 
equal logarithmic scales for three values of $z$ roughly in the middle of each 
interval $z\in[\mu_j,\mu_{j+1}]$ (top). The zero-mode with anti-periodic 
boundary conditions is found at $z=30/60$.}\label{fig:7}
\end{figure}
As discussed earlier in this section, 
each constituent monopole is associated with an SU(2) embedding and the two 
isospin components of these embeddings effectively have ``masses" $2\pi(\mu_m
/\beta-z)$ and $2\pi(\mu_{m+1}/\beta-z)$. This only gives a normalizable 
zero-mode when these are of opposite sign~\cite{Call}, that is for $z\in[
\mu_m/\beta,\mu_{m+1}/\beta]$. In the interior of this interval the zero-mode 
is exponentially localized (to the constituent monopole at $\vec y_m$), but at 
$z=\mu_m$ or $z=\mu_{m+1}$ one of the isospin components becomes massless and 
the zero-mode delocalizes, having in addition support on the constituent at 
respectively $\vec y_{m-1}$ or $\vec y_{m+1}$.

The expression for the zero-mode density can be given in a simple
form for arbitrary charge 1 calorons as well~\cite{ZMN}. With 
$\mu_m/\beta\!\leq\!z'\!\leq\!z\!\leq\!\mu_{m+1}/\beta$ 
\beqa
&&\hskip-3mm\hat f_x(z,z')=\frac{\pi e^{2\pi i t(z-z')}}{\beta r_m\psi}\langle 
v_m(z')|\cA_{m\!-\!1}\cdots \cA_1\cA_n\cdots \cA_m-e^{-2\pi it/\beta}|\sigma_2
v_m(z)\rangle,\nonumber\\&&\hskip-3mm v_m(z)\equiv\pmatrix{\sinh[2\pi(z-
\mu_m/\beta)r_m]\cr\cosh[2\pi(z-\mu_m/\beta)r_m]\cr},\quad|\hat\Psi_z(x)|^2
=-\frac{1}{4\pi^2}\partial_\mu^2\hat f_x(z,z),\label{eq:green}
\eeqa
where for later use we have introduced also the off-diagonal expression for 
the Green's function $\hat f_x(z,z')$ (for $z\!\leq\!z'$ one can use the 
fact that $\hat f_x(z',z)=\hat f_x^*(z,z')$), which can be computed using 
impurity scattering calculations in a piecewise constant potential, see 
Sec.~\ref{sec:expl}, the details of which need not concern us here. We 
illustrate the behavior of the zero-mode in Fig.~\ref{fig:7} at various 
values of $z$ for the caloron of Fig.~\ref{fig:5}. It is also interesting to 
consider the so-called zero-mode limit (zm) in which $\vec x$ is far 
removed from any constituent location, except the one at $\vec y_m$. This
gives\footnote{Allowing only for errors decaying {\em exponentially} in 
$r_{l\neq m}$, one needs to include $\cO(r^{-1}_{m\pm1})$ dependent shifts 
in $\mu_m$, $\mu_{m+1}$ and $\nu_m$, see \refeq{zmlim} and Ref.~\cite{BNvB}.} 
\beq
\hat f_x^\zm(z,z)=\frac{2\pi\sinh\left(2\pi r_m(\mu_{m+1}/\beta-z)\right)
\sinh\left(2\pi r_m(z-\mu_m/\beta)\right)}{\beta r_m\sinh\left(2\pi r_m
\nu_m/\beta\right)}.
\eeq
At $z=\half(\mu_m+\mu_{m+1})/\beta$ (i.e. $z=0$ or $z=1/2\beta$ for SU(2)) 
we find 
\beq
\hat f_x^\zm(z,z)=\pi(r_m\beta)^{-1}\tanh(\pi r_m\nu_m/\beta),
\eeq 
with $-(4\pi^2)^{-1}\partial_\mu^2\hat f^\zm_x(z,z)$ giving precisely the 
zero-mode density of a basic BPS monopole, confirming once again the nature 
of the constituents. In the high temperature limit, as long as $z\neq\mu_m/
\beta$, the zero-modes become infinitely localized to the constituent 
locations $\vec y_m$. In this limit the zero-mode density is given by 
$\beta^{-1}\delta(\vec x-\vec y_m)$ for $\mu_m/\beta\!<\!z\!<\!\mu_{m+1}/
\beta$. The zero-modes are therefore ideal for localizing the cores of the 
constituent monopoles, particularly useful for the higher charge calorons 
discussed below. It should be noted that in general the definitions of the 
constituent locations based on the peaks in the energy density, the degeneracy 
of two eigenvalues in the Polyakov loop and the peak in the zero-mode 
densities will only coincide with $\vec y_m$ when all constituents have a 
non-zero mass and are well separated (as compared to the size of the monopole 
cores) from each other.

Before discussing the higher charge calorons we give the simple expression 
for the SU(2) charge 1 gauge field in the algebraic gauge with 
$\pl=\exp(\beta\omega\tau_3)$ (hence $\mu_2=-\mu_1=\beta\omega$),
\beq
A_\mu=\frac{i}{2}\bar\eta^3_{\mu\nu}\tau_3\partial_\nu\log\phi+
\frac{i}{2}\phi{\rm Re}\left((\bar\eta^1_{\mu\nu}-i\bar\eta^2_{\mu
\nu})(\tau_1+i\tau_2)\partial_\nu\chi\right),\label{eq:ag}
\eeq
where $\phi^{-1}\equiv 1-\rho^2\hat f_x(\omega,\omega)$ and $\chi\equiv
\rho^2\hat f_x(\omega,-\omega)$, which has some resemblance to the 't~Hooft 
ansatz in \refeq{tHooft}. It reduces to this ansatz for trivial holonomy, 
$\omega=0$, for which $\chi=1-\phi^{-1}$, and one easily checks that this 
gives precisely the Harrington-Shepard solution with $\phi=\phi_{HS}$, see
\refeq{HS}. In the high temperature limit one finds for non-trivial holonomy 
$\chi^\ff=0$, $\phi^\ff=(|\vec x-\vec y_1|+|\vec x-\vec y_2|+|\vec y_2-\vec 
y_1|)/(|\vec x-\vec y_1|+|\vec x-\vec y_2|-|\vec y_2-\vec y_1|)$ and that 
$A_\mu=\frac{i}{2}\bar\eta^3_{\mu\nu}\tau_3\partial_\nu\log
\phi^\ff$ describes a pair of oppositely charged Dirac monopoles with the 
Dirac string on the line connecting them, where $\phi^\ff$ diverges, but 
outside of which $\log\phi^\ff$ is harmonic. 

\subsection{Higher charge calorons}\label{sec:higher}
When ignoring charged components of the gauge field, outside the core the 
Abelian field has unavoidable Dirac strings. We can trace how the exact 
solution instead takes care of the return flux, namely through the Abelian 
component in the magnetic field coming from the commutator of the charged 
components of the non-Abelian field, as is of course well-known from the 
't~Hooft-Polyakov monopole. 
\begin{figure}[htb]
\vskip4.1cm
\includegraphics{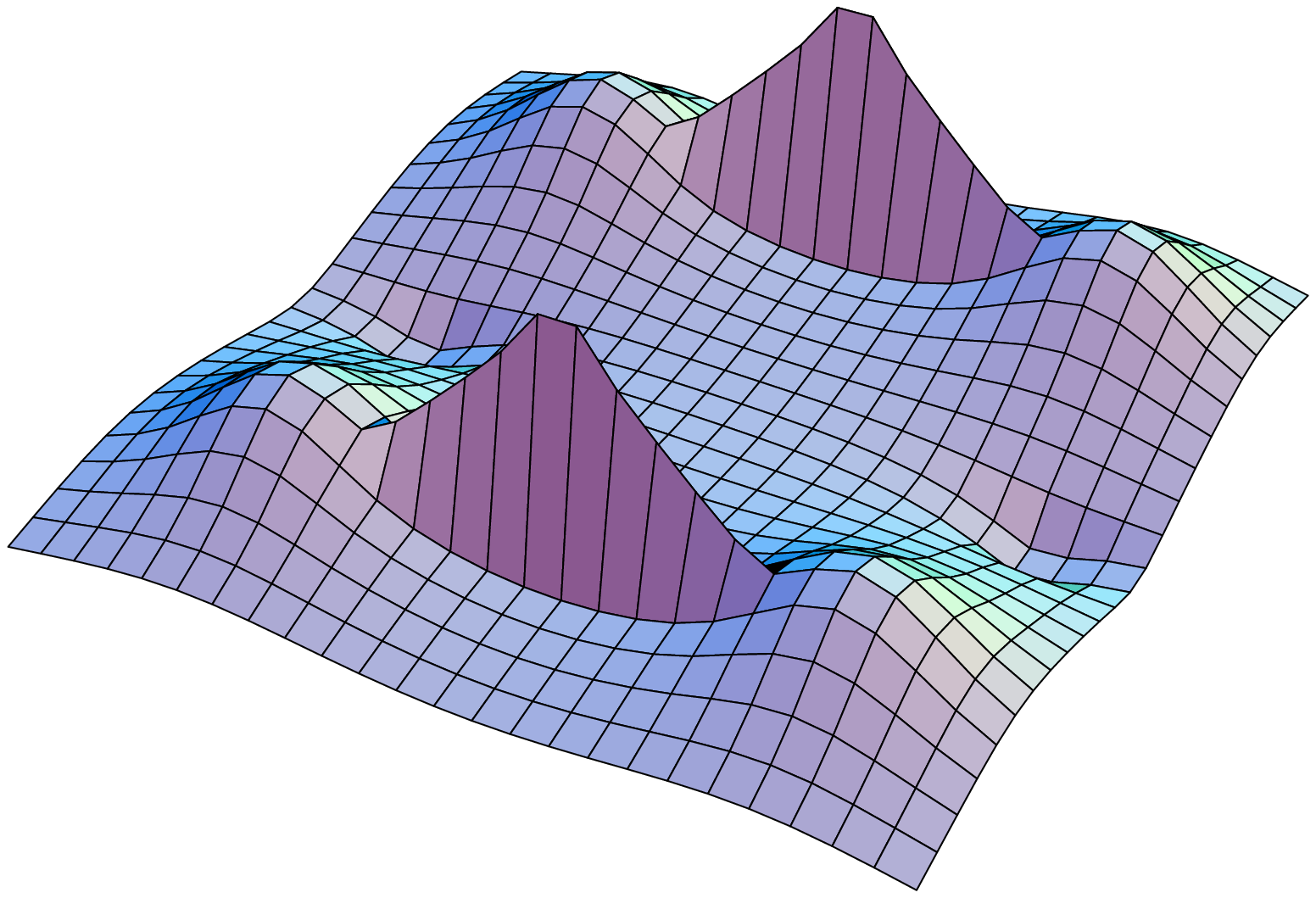}
\includegraphics{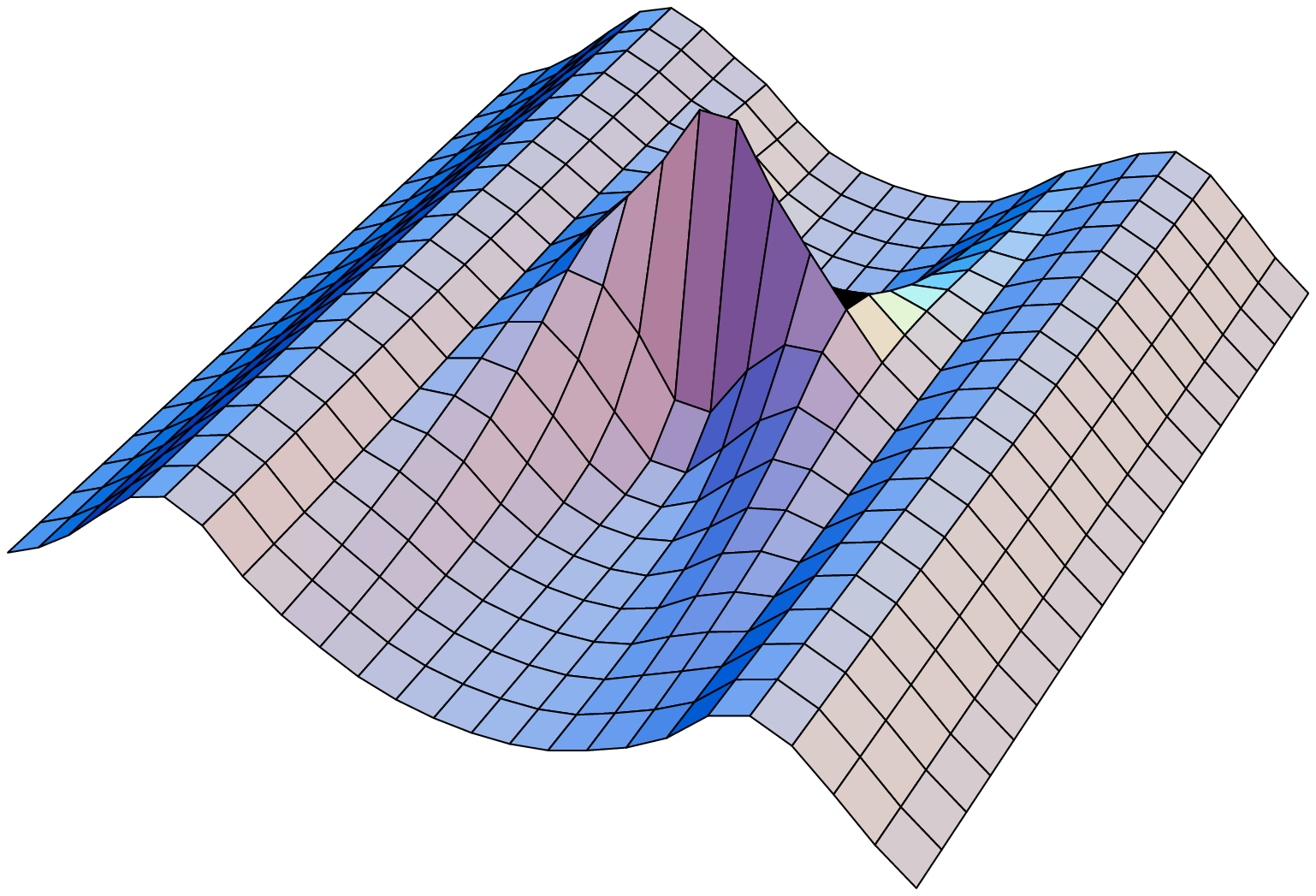}
\caption{Approximate superposition of two $SU(2)$ charge 1 calorons (left). 
The logarithm of the action density is plotted through the plane of the 
constituents at $t=0$ for $\beta=1$ we zoom in by a factor 40 for the 
transverse direction on the would-be Dirac strings (right).}\label{fig:8}
\end{figure}
But from the numerical point of view this will 
require exponential fine tuning outside the core. It is notoriously difficult 
to find approximate superpositions for magnetic monopoles without seeing a 
remnant of the Dirac string, and in that sense there is ``no free lunch". Like 
in the instanton liquid we would like to make approximate superpositions of 
calorons, which also allows us to mix calorons of different charges. However, 
for those that ``dissociate", the constituents should not ``remember" from 
which caloron they originated. Although the constituents themselves might be 
well separated, interference can occur in the regions between them. An example 
of this is shown in Fig.~\ref{fig:8}, where we added two charge 1 caloron 
gauge fields in the algebraic gauge, called the sum ansatz~\cite{ScSh,Diak}. 
This preserves the gauge condition but some care is required at the gauge 
singularity, which is removed by a gauge transformation with non-trivial 
winding number (sometimes one refers to these gauge fields as being in the
singular gauge). Adding the gauge fields after such a gauge transformation 
is performed, destroys the proper decay at infinity. Instead, one first
smoothly deforms to zero the gauge fields of all other calorons in a small 
neighborhood centered around the gauge singularity one wishes to remove.
This only costs a small amount of action, particularly when the calorons are 
not too close. In principle a similar construction is possible for keeping 
Dirac strings hidden, but in this case the gauge singularity is due to 
approximating the fields by their Abelian component, far from the constituent 
cores. However, this would require exponential fine tuning. Nevertheless, not 
performing any adjustment the action density along the Dirac string (or 
sheet due to the additional extent in the time direction) actually stays 
finite\footnote{Only at the gauge singularity the action density of the 
combined field diverges, which may be removed as discussed.}, see 
Fig.~\ref{fig:8}.

\begin{figure}[htb]
\vskip2.9cm
\includegraphics{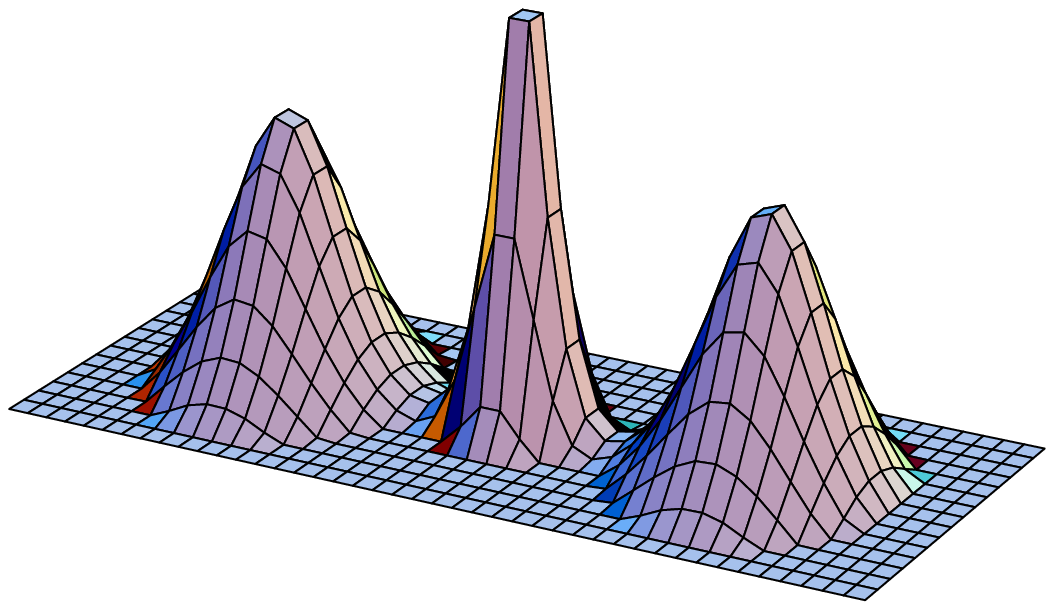}
\includegraphics{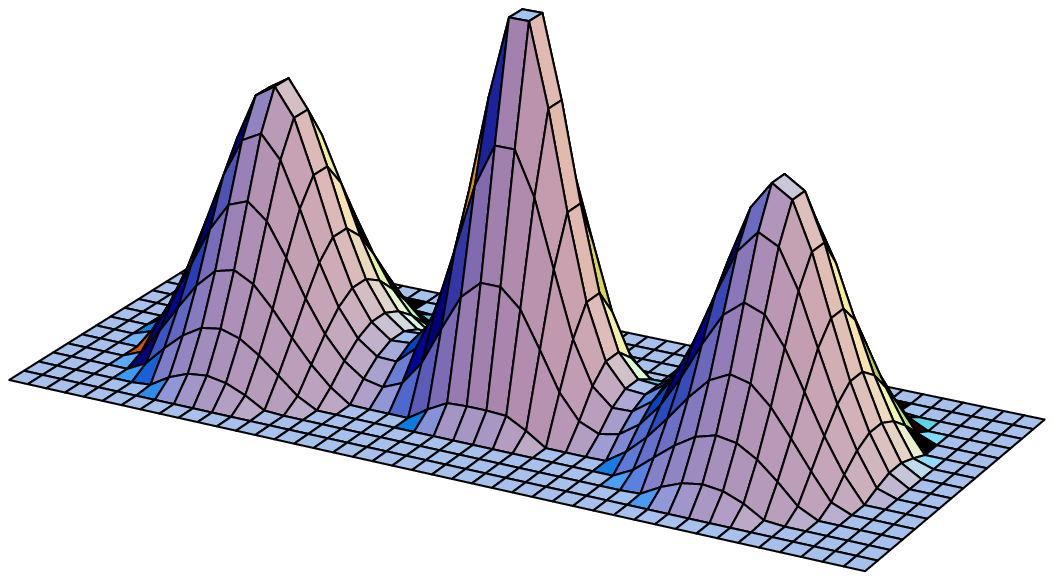}
\includegraphics{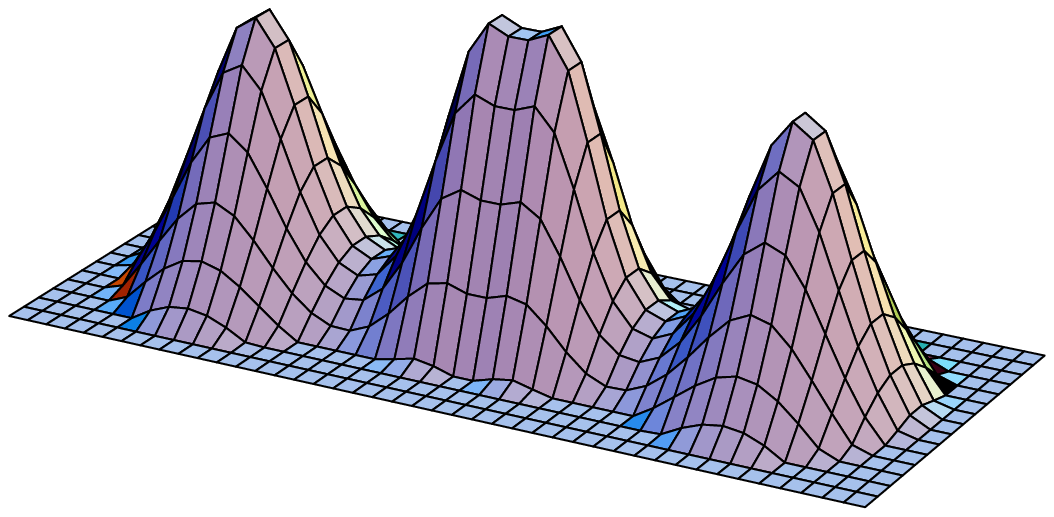}
\caption{
Action density on equal logarithmic scales for exact axially 
symmetric SU(2) charge 2 solutions at $\beta=1$ for $t=0$, on a plane 
through the constituents. Locations can be read off from the dotted 
lines in Fig.~\ref{fig:4}, from left to right. For the last case see 
Fig.~\ref{fig:10}.}\label{fig:9}
\end{figure}

\begin{figure}[htb]
\vskip3.2cm
\includegraphics{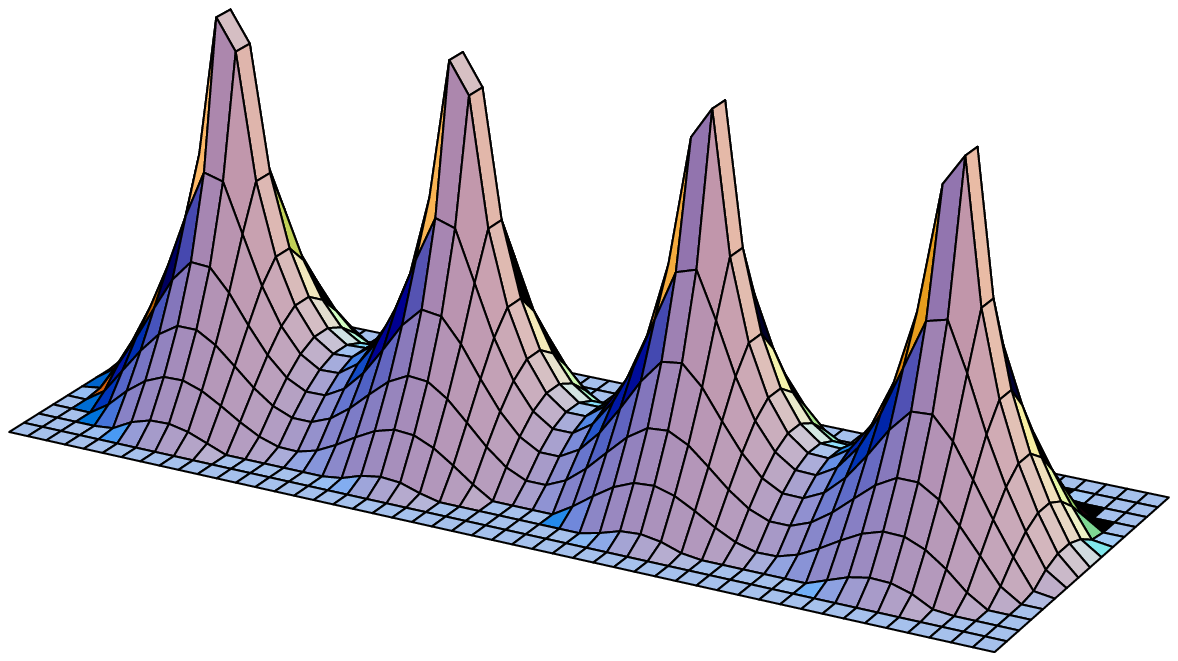}  
\includegraphics{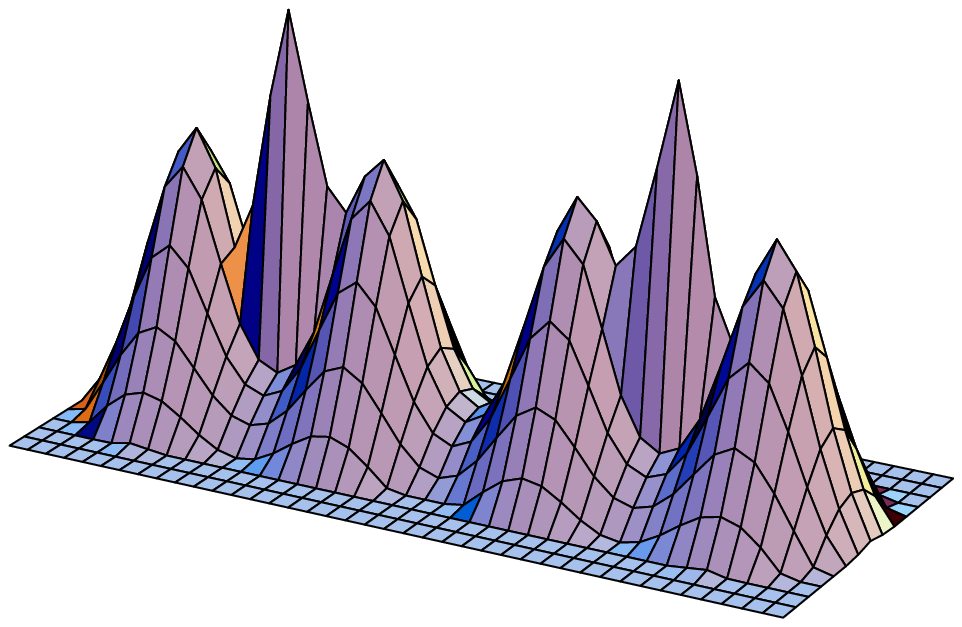}
\includegraphics{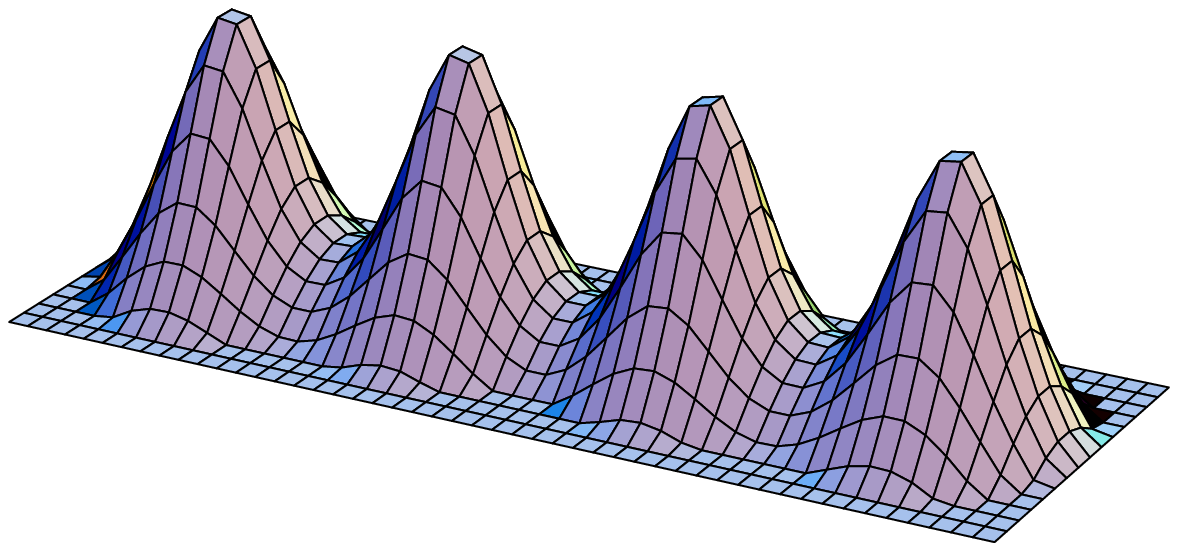}
\caption{Action density on equal logarithmic scales for the exact axially 
symmetric SU(2) charge 2 solution with all constituents well-separated (left), 
the approximate solution based on the sum ansatz (middle) and the high 
temperature limit with point-like Dirac monopoles (right). See also
Fig.~\ref{fig:9}.}\label{fig:10}
\vspace{3.4cm}
\includegraphics{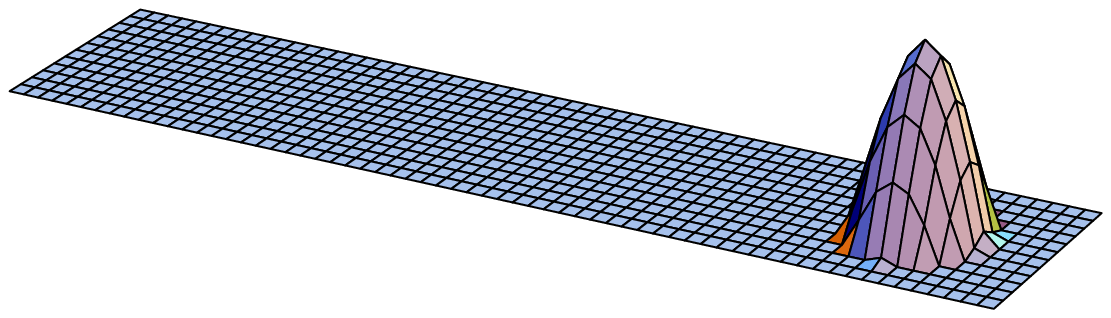}
\includegraphics{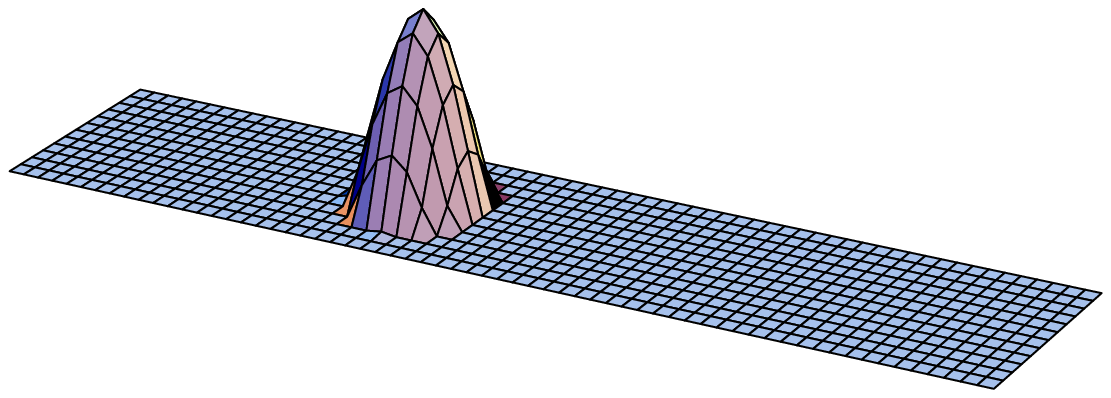}
\includegraphics{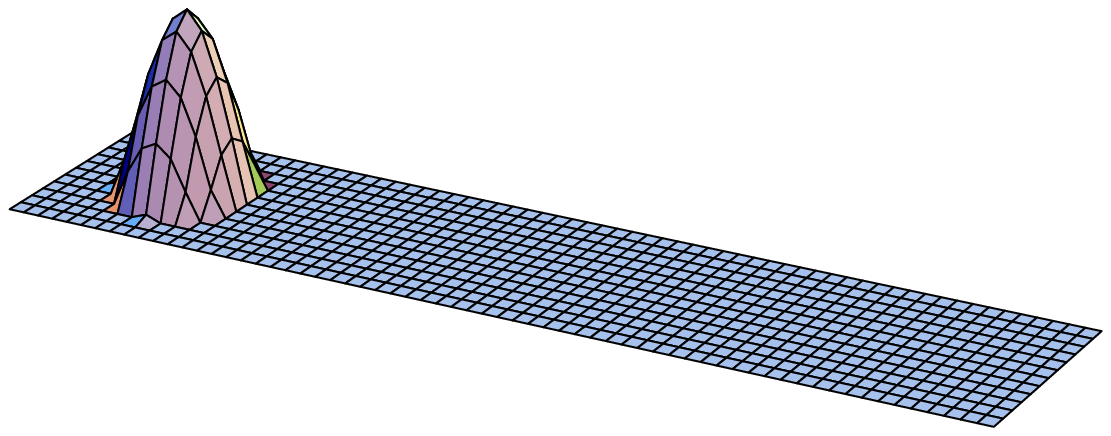}
\includegraphics{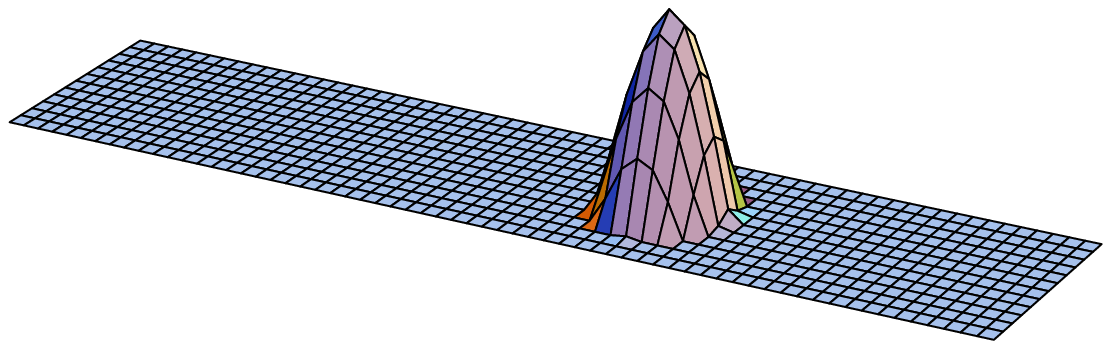}
\caption{Zero-mode densities on equal logarithmic scales for the caloron in
Fig.~\ref{fig:10} (same horizontal scale). On the left are shown the two 
periodic zero-modes ($z=0$) and on the right the two anti-periodic zero-modes 
($z=1/2$).}\label{fig:11}
\end{figure}

It is due to these complications we felt compelled to analyse the higher charge
caloron solutions. The disadvantage is that one can only consider the exact 
self-dual solutions this way, and not superpositions of opposite charges. 
On the other hand, it should guarantee absence of Dirac strings, as we indeed 
confirm for a set of exact axially symmetric solutions that share all the 
properties of the charge 1 solutions, as illustrated for the example of SU(2) 
charge 2 solutions in Figs.~\ref{fig:9}-\ref{fig:11}. In the high temperature 
limit these solutions are again described by point-like constituent 
monopoles~\cite{FBvB}. 
Most of the memory effect has disappeared. 
In the light of this it is interesting to point out that when further 
separating the two pairs on the left from those on the right, each can be 
viewed as coming from a single caloron\footnote{In this context the parameters 
$\xi_a$ and $\rho_a$ appearing in \refeq{Ym} can be interpreted as the center 
of mass and size of those calorons.}. However, when the two middle constituents
start to get closer than the size of their cores, as in Fig.~\ref{fig:9} it 
is more natural to interpret the solution in terms of one well ``dissociated" 
caloron formed by the two outer constituents, on top of which is superposed a 
small ``non-dissociated" caloron, which is no longer static. Indeed, its peak 
is nearly $O(4)$ symmetric, whereas the other two constituents are static and 
$O(3)$ symmetric to a high degree, as is appropriate for a BPS monopole (see 
Fig.~\ref{fig:9} (left)). 
A very subtle memory effect, however, remains. It can be shown that for these 
point-like axially symmetric solutions the magnetic charges have to alternate. 
In particular one cannot move one monopole through the other. It may perhaps 
come as a surprise, but in part there is a good reason for being cautious, 
since as we have just seen two oppositely charged constituent monopoles really 
form a small caloron.  Since the distance between constituents is given by 
$\pi\rho^2/\beta$, this means that we have to go through a singular caloron 
when interchanging constituent locations on the line. A singular caloron lies 
on the boundary of the moduli (i.e. parameter) space and one cannot use 
continuity arguments.

The reason to expect that in general one no longer deals exclusively with 
point-like monopoles comes from the known multi-monopole solutions. It is 
well-known that when putting monopoles of equal charge on top of each other, 
these are deformed in for example the shape of a doughnut~\cite{Ring}, as 
famously illustrated in the scattering of two magnetic monopoles~\cite{AtHi}. 
At the technical level this is related to the fact that the Green's function 
mentioned above no longer involves a piecewise constant potential. Only for
the axially symmetric solutions this was still the case. Nevertheless, for 
SU(2) and charge 2 we have been able to find an exact expression for the 
zero-mode density in the high temperature limit. As long as $z\neq\mu_j$, the 
fermions have an infinite mass in this limit, and the zero-modes will vanish 
outside the cores of the constituent monopoles, which themselves need not 
necessarily be isolated points. We thus are able to trace with the help of 
the zero-modes to which region the cores have to be localized. Here we will 
just present the result~\cite{BNvB} and discuss its physical significance. For
$\mu_m\!<\!z\!<\!\mu_{m+1}$ and in the far field limit $\sum_a|\hat\Psi_z^a
(x)|^2=-\beta^{-1}\partial_i^2\cV_m(\vec x)$, with
\beq
\cV_m(\vec x)=\frac{1}{2\pi|\vec x|}+\frac{\cD}{4\pi^2}\int_{r<\cD}\hskip-5mm
drd\varphi~\frac{\partial_r|\vec x-r\vec y(\varphi)|^{-1}}{\sqrt{\cD^2-r^2}},
\eeq
where $\vec y(\varphi)\!=\!(\sqrt{1-\k^2}\cos\varphi,0,\sin\varphi)$, up to an
arbitrary coordinate shift and rotation. Here $\cD$ is a scale and $\k$ a shape 
parameter to characterize {\em arbitrary} SU(2) charge 2 solutions. In this 
representation it is clear that $\cV_m(\vec x)$ is harmonic everywhere except 
on a disk bounded by an ellipse with minor axes $2\cD\sqrt{1\!-\!\k^2}$ and 
major axes $2\cD$. Although not directly obvious, when $\k\to1$ the support of 
the singularity structure is on two points only, separated by a distance 
$2\cD$. Taking an arbitrary test function $f(\vec x)$ one can prove 
that~\cite{BNvB} 
\beq
\lim_{\k\to1}-\int f(\vec x)\partial_i^2\cV_m(\vec x)d^3x=f(0,0,\cD)+
f(0,0,-\cD).
\eeq

\begin{figure}[htb]
\vskip1.3cm$~\k$\vskip3.2cm\hskip5.7cm$\rho$\vskip3mm
\includegraphics{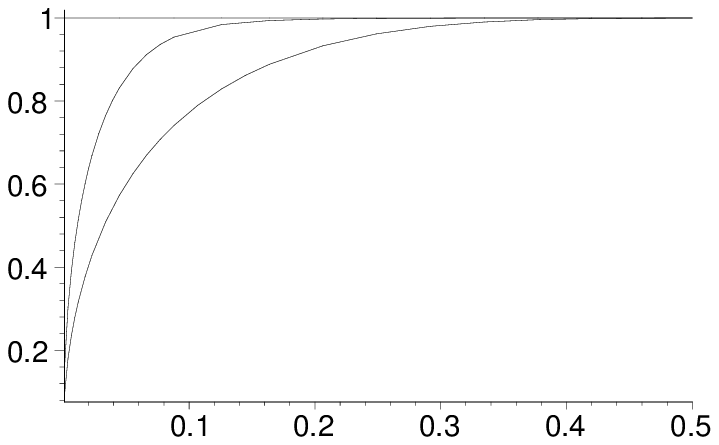}
\includegraphics{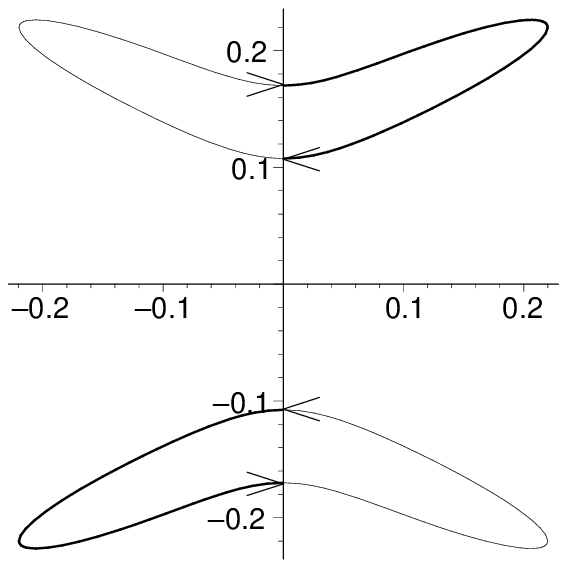}
\caption{On the left we plot $\k$ versus $\rho$ for $\alpha=0$ ($\k\equiv1$),
$\alpha=-\pi/100$ and $\alpha=-\pi/2$. On the right are shown the locations 
of the constituent monopoles (fat vs. thin curves for opposite charges) at 
$\rho=1/4$, varying $\alpha$ from $-\pi$ (indicated by the arrows) to 
0.}\label{fig:12}
\end{figure}

For the caloron $\k$ and $\cD$ are in general not independent, monopoles of 
different charges have to adjust to each other to form an exact caloron 
solution. As an example we illustrate in Fig.~\ref{fig:12} the relation 
for the case studied in Ref.~\cite{BNvB},
which is a two parameter family of exact SU(2) charge 2 calorons solutions 
in terms of an instanton scale $\rho$ and relative gauge orientation angle 
$\alpha$. For fixed $\rho$ it interpolates between two axially symmetric 
solutions. On the left is shown the relation between $\k$ and $\rho$ for 
some values of $\alpha$. Together with the fact that $\cD=\pi\sqrt{2}
\rho^2(1+\cO(1-\k^2))$, this shows that $\k\to1$ for increasing $\cD$. 
It can be shown this approach is exponential, cmp.~Fig.~\ref{fig:12} (left).

The important conclusion is that, when well separated, the constituent 
monopoles become point-like objects. 
This is a necessary requirement for 
the constituent monopoles to be used as entities to describe the field 
configurations at larger distances. Much work of course remains to be 
done, but the results so far have been encouraging. In addition important 
lattice evidence has been accumulated by now, which we will briefly 
summarize in the next section.

\section{Lattice evidence}\label{sec:latt}
There are a number of lattice studies, that clearly point to constituent 
monopoles to have a dynamical role to play. Two methods have been employed. 
One is based on so-called cooling, the other one uses fermion zero-modes. 
In both cases the purpose is to filter out ultraviolet noise, i.e. one is 
interested in the long distance fluctuations. 

\subsection{Cooling}\label{sec:cool}
Cooling is the process by which one lowers the lattice action through local 
updates, replacing a link by a suitable combination of neighboring links, 
such that when remaining unchanged it satisfies the lattice equations of 
motion. There are by now many variants, using improved lattice actions (to 
reduce discretization errors), and criteria to stop the cooling~\cite{Owe}. 
Such a criterion is important, because when cooling too long either all the 
non-trivial fields are removed, or at best one reaches a self-dual solution. 
The latter is of course sometimes done on purpose, so as to reproduce the 
classical solutions on the lattice to quite some precision, or to look for 
solutions not known exactly. For the calorons this has been extensively 
studied~\cite{CCal}, even in connection with a numerical implementation of 
the Nahm transformation~\cite{TNahm}, mainly in the presence of so-called 
twisted boundary conditions~\cite{Twist} to coach the system in having 
non-trivial holonomy. 

{}From the dynamical point of view, one would like to find how often, and with 
which properties, calorons appear in the long distance fluctuations. In this 
case one would like to start from Monte Carlo generated configurations and 
stop the cooling process when reaching a plateau in the plot for the action 
as a function of the number of cooling steps. The plateau would be infinitely 
long if the configuration is a solution of the lattice equations of motion, 
but would still be sizeable if these are satisfied approximately. Therefore, 
configurations consisting of any number of well-separated instantons ($Q$) and 
anti-instantons ($\bar Q$) can form such plateaus, with an action approximately
equal to $8\pi^2(Q+\bar Q)$, whereas the topological charge of the equivalent 
continuum configuration equals $Q-\bar Q$. An instanton can shrink to such a 
small size $\rho$ (from the caloron point of view its two constituents 
monolopes getting too close together) that the lattice no longer supports it
as a solution\footnote{This can be prevented by a suitable choice of improved 
lattice action~\cite{Snip}.}. Also, when instantons and anti-instantons come 
close together they annihilate, this effect is independent of the lattice 
discretization.  In both cases the plateau ends, and the cooling curve 
typically settles down at the next plateau where $Q+\bar Q$ has decreased 
by 1 or 2 unit respectively (whereas $Q-\bar Q$ changes by $\pm1$ or 0
respectively). The expectation is that the annihilation of oppositely charged 
configurations is slow, at least when they are sufficiently separated, and 
that from the plateaus one can recover the statistical properties of 
instantons and calorons. 

\begin{figure}[htb]
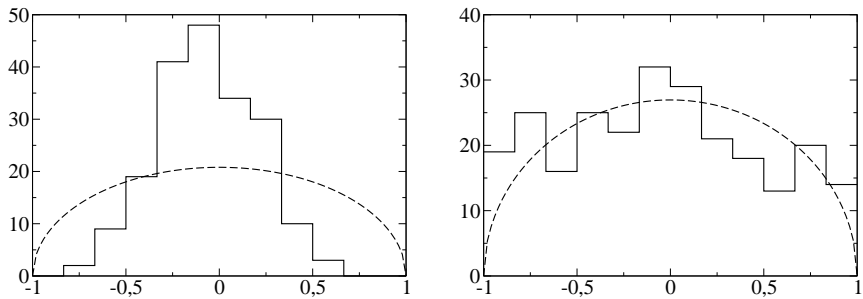

\vskip4.7cm
\includegraphics{hol.eps}
\includegraphics{hol1.eps}
\caption{The ``asymptotic holonomy" $L_\infty$ for an ensemble of $\cO(200)$ 
configurations generated on a $16^3\times 4$ lattice with periodic boundary 
conditions for a lattice coupling $4/e^2=2.2$. Left is shown the measurement 
on the first plateau, right on the last plateau. For comparison the Haar 
measure $\sqrt{1-L^2}$ is shown by the dashed curve. Figures taken from 
Ref.~\cite{IMMV}.}\label{fig:13}
\end{figure}

To prompt the system to have a given holonomy, one can freeze the time-like 
links at the spatial boundary of the box to the required holonomy~\cite{FixL}, 
but in larger volumes it is expected that the confining environment itself will 
provide local regions with a sufficiently coherent background $A_0$ field 
associated with non-trivial holonomy. One important finding has been that after 
cooling the non-trivial holonomy is preserved to some degree~\cite{IMMV}. This 
has been analysed for SU(2) in terms of the so-called ``asymptotic holonomy" 
$L_\infty$, defined as the average of $\half\Tr P(\vec x)$ over all points 
$\vec x$ for which the action density, summed over $t$, is smaller than .0001. 
Histograms of $L_\infty$ in the confined phase are shown in Fig.~\ref{fig:13} 
(taken from Ref.~\cite{IMMV}). Early on in the cooling process a clear peak at 
$L_{\infty}=0$ is observed, which becomes a flat distribution when cooled down 
to the plateau associated to $Q+\bar Q=1$. Important is in particular that the 
distribution is {\em not} becoming peaked towards $L_\infty=\pm1$. Many 
configurations with well ``dissociated" calorons in this dynamical setting have
been found~\cite{IMMV}, and with this method one can study the configurations 
in great detail. For SU(2) one can thus look for the monopole centers by 
finding where the Polyakov loop equals $\pm\Eins_2$ and use the fermion 
zero-modes for periodic and anti-periodic boundary conditions to test if 
they are localized on the appropriate constituents\footnote{For SU(2) one 
profits from the fact that $z=0$ always lies in the middle of the interval 
$[\mu_1,\mu_2]$, whereas $z=\half$ always lies in the middle of 
$[\mu_2,1+\mu_1]$.}. 

In these studies one also finds configurations that cannot be directly 
interpreted in terms of instantons, calorons and their possible constituent
monopoles~\cite{IMMV}. Interestingly, there are cases where two constituents
appear, but of opposite fractional topological charge. These could arise from 
the ``annihilation" of two other constituents of opposite duality~\cite{Anni}
but where originally each of these, together with one of the surviving 
constituents, formed an (anti-)caloron. More recently these authors also 
performed cooling studies for SU(2) at considerably lower temperature 
than just around the deconfining phase transition~\cite{Ilg}. The calorons 
in this case do not tend to ``dissociate" in isolated lumps of action density. 
Nevertheless, constituents could be identified through the behavior of the 
Polyakov loop and were characteristic for non-trivial holonomy. This may 
explain why in the past constituent monopoles were never noticed in cooling 
studies. Finally let us mention that many results for SU(3) have now been 
obtained as well~\cite{Ilg}. 

\subsection{Zero-mode filter}\label{sec:zmfilter}
\begin{figure}[htb]
\vskip9.3cm
\includegraphics{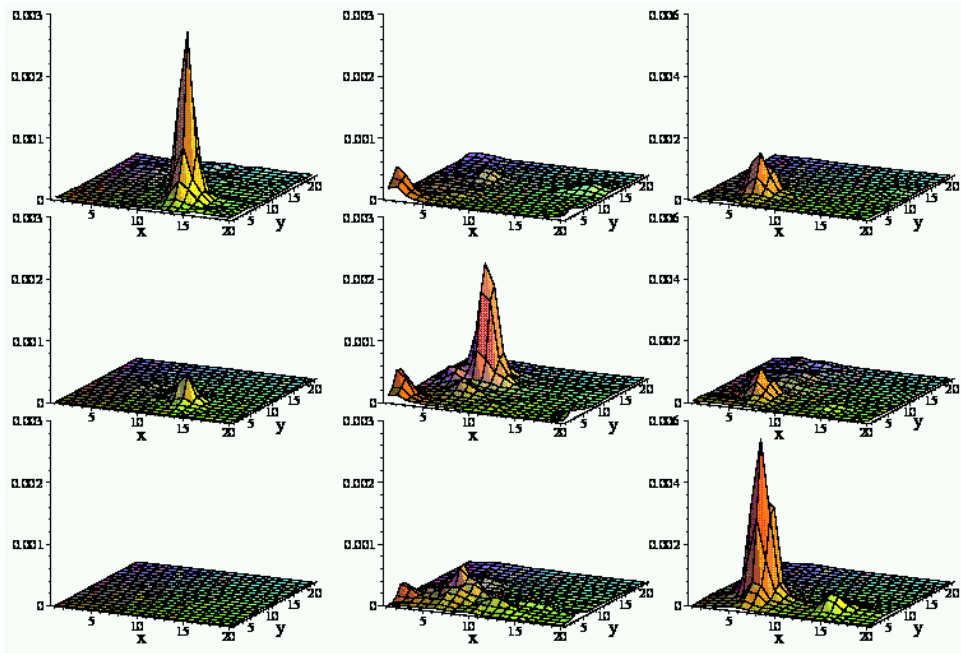}
\caption{The zero-mode density for a particular configuration in the 
confined phase on a $6\times20^3$ lattice with (L\"uscher-Weisz) coupling 
$6/e^2=8.20$, taken from Ref.~\cite{GaSc}. Shown are slices in $(x,y)$ at 
$(t,z)=(5,9)$ (left), $(t,z)=(2,19)$ (center) and $(t,z)=(5,18)$ (right), 
for $z=0.05$, 0.3 and 0.65 (from top to bottom).}\label{fig:14}
\end{figure}

The use of zero-modes as a filter relies on two observations. The first is, 
as we have seen, that zero-modes quite accurately trace the underlying gauge 
field of instantons, calorons and constituent monopoles. Secondly, a zero-mode 
probes the long distance features of the configuration and ignores the 
ultraviolet high momentum components, otherwise one could never have a 
zero-mode. In some sense the Dirac operator is a particular projection of the 
covariant momentum. For SU(3), comparing periodic and anti-periodic fermion 
boundary conditions~\cite{Gatt}, or cycling through all possible 
phases~\cite{GaSc}, \refeq{cycle}, a significantly different behavior in 
the two phases was found. In the deconfined phase, the proper $z$-dependent 
behavior of the trivial holonomy configuration is seen. These are of course 
the old Harrington-Shepard solutions, but their zero-modes were previously 
only considered for anti-periodic boundary conditions. In the confined phase 
indications for a three lump structure is seen (see Fig.~\ref{fig:14}, taken 
from Ref.~\cite{GaSc}) in a reasonable fraction of the configurations.
These results are based on dynamical configurations generated with the 
L\"uscher-Weisz~\cite{LuWe} improved lattice action. Configurations that 
had exactly one chiral zero-mode (which requires the use of a chirally 
improved lattice Dirac operator~\cite{ChIm}) were singled out. By the index 
theorem, this implies the topological charge is equal to one. The dynamically 
generated configurations can, and typically will, consist of Q+1 instantons and 
Q anti-instantons, giving rise to near zero-modes with less perfect chiral 
behavior. Work is in progress to analyse the near zero-modes, as well as 
performing cooling on the SU(3) gauge field configurations~\cite{Comm}. This 
will help to further rule out the unlikely possibility for the zero-mode to 
jump between instanton, rather than monopole constituents. A more theoretical 
study~\cite{Tsuk} has recently shown that it seems indeed impossible for
the full signature of a single caloron with non-trivial holonomy to be 
emulated by the effect of hopping between instantons.

Even more remarkable is that a similar structure of multiple lumps for the 
topological charge 1 sector was found on a symmetric lattice of size $16^4$ 
and $12^4$ at low temperature, well in the confined phase~\cite{GaSc,GatL}. In 
this case the lattice does not single out one direction as being (imaginary)
time. Yet, using one of these directions to impose the $z$ dependent boundary 
condition for the fermions, in almost half of the configurations one finds 
more than one lump. The results seem to indicate these lumps are randomly 
distributed over the volume. How to reconcile this with the results found 
with cooling, that pointed to ``non-dissociated" calorons at low temperature, 
is at the moment not clear. But let us recall that fractionally charged 
instantons on the torus were long ago introduced by 't~Hooft~\cite{Twist}, 
and have been extensively studied on the lattice~\cite{GaGo}. Twisted 
boundary conditions and, related to it, the quantization of the topological 
charge in units of $1/n$, make it more difficult to embed these solutions 
in a dynamical environment. Nevertheless, on the basis of some simple 
assumptions concerning the dynamical properties of these configurations, 
quasi realistic results have been obtained~\cite{MaGo}. It is indeed 
compelling to interpret the $4kn$ dimensions of the SU($n$) charge $k$ 
instanton parameter space on the torus in terms of the space-time locations 
of $kn$ instanton quarks, as they were called long ago~\cite{Inqu} (even 
though their meaning at that time was more abstract). Results for 
instantons on $T^2\times R^2$, that in some sense interpolate between the 
case of calorons and instantons on the torus, give further 
evidence~\cite{Ford} for this conjecture.

\section{Conclusion}\label{sec:concl}
It is clear we can do no justice here to all the results that have been 
obtained in using lattice simulations. The great advantage of using 
zero-modes as a filter is that it minimizes the bias, since it directly uses 
the Monte Carlo configurations themselves. On the other hand, in the cooling 
studies one has access to the relation between the behavior of the Polyakov 
loop, which plays the role of an order parameter in SU($n$) gauge theories, 
and the presence of constituent monopoles, which may give some insight in 
the underlying dynamics. Of course, we would like to make constituent 
monopoles into a precise tool, e.g. for testing the celebrated idea of a dual 
superconductor to describe confinement~\cite{Mand,Tho5}, or as we speculated 
in the introduction, to describe a dense phase of monopoles to lead to 
deconfinement of magnetic charges, like quark deconfiment at high density. 

On the more theoretical side we have also not touched upon the role these 
constituent monopoles play in supersymmetric gauge theories~\cite{DHKM,DiPe} 
and how the initial motivation for reconsidering calorons with non-trivial 
holonomy came from the study of D-branes~\cite{LeYi,PLB,LeYi2}. Nevertheless, 
we feel the applications to the dynamics of non-Abelian gauge theories is very 
promising. We hope to have given the reader some insight in this matter.

\section*{Acknowledgements}
PvB would like to say to the organizers, and Michal Praszalowicz in particular, 
dzi\c ekuj\c e bardzo for inviting him to Zakopane again, this time 
coinciding with the historic moment of the referendum to join the European 
Union. He also thanks Maxim Chernodub, Margarita Garc\'{\i}a P\'erez, Tony 
Gonz\'alez-Arroyo, Thomas Kraan, Alvaro Montero and Carlos Pena for the 
many fruitful collaborations over the last 5 years, and Christof Gattringer, 
Michael Ilgenfritz, Boris Martemyanov and Michael M\"uller-Preussker for 
inspiring discussions concerning their lattice studies. The research of FB 
is supported by FOM.

\end{document}